\documentclass[twocolumn,prc,floatfix,showpacs,preprintnumbers,nofootinbib,%
superscriptaddress]{revtex4}
\usepackage[utf8x]{inputenc}
\usepackage{mathrsfs}
\usepackage{amssymb}
\usepackage{amsmath}
\usepackage{graphicx}

\usepackage{xcolor}
\definecolor{lcolor}{rgb}{0.5,0,0}
\definecolor{citcolor}{rgb}{0,0.3,0.0}

\usepackage[breaklinks,colorlinks,urlcolor=blue,citecolor=citcolor,linkcolor=lcolor]{hyperref}
\usepackage{mciteplus}

\newcommand{\der}{\mathrm{d}}
\newcommand{\rt}{{\mathbf{r}_T}}

\newcommand{\bt}{{\mathbf{b}_T}}
\newcommand{\st}{{\mathbf{s}_T}}

\newcommand{\pt}{{\mathbf{p}_T}}
\newcommand{\qt}{{\mathbf{q}_T}}
\newcommand{\kt}{{\mathbf{k}_T}}

\newcommand{\cf}{C_\mathrm{F}}

\newcommand{\nr}[1]{(\ref{#1})}

\newcommand{\qs}{Q_\mathrm{s}}

\newcommand{\qso}{Q_\mathrm{s0}}

\newcommand{\lqcd}{\Lambda_{\mathrm{QCD}}}
\newcommand{\as}{\alpha_{\mathrm{s}}}
\newcommand{\aem}{\alpha_{\mathrm{em}}}

\newcommand{\eq}{Eq.~}
\newcommand{\se}{Sec.~}

\newcommand{\Ncal}{{\mathcal{N}}}

\begin{document}

\author{T. Lappi}
\affiliation{
Department of Physics, %
 P.O. Box 35, 40014 University of Jyväskylä, Finland
}

\affiliation{
Helsinki Institute of Physics, P.O. Box 64, 00014 University of Helsinki,
Finland
}
\author{H. Mäntysaari}
\affiliation{
Department of Physics, %
 P.O. Box 35, 40014 University of Jyväskylä, Finland
}

\title{

Single inclusive particle production at high energy from
HERA data to proton-nucleus collisions.
}


\preprint{}

\begin{abstract}
We study single inclusive hadron production in proton-proton and 
proton-nucleus collisions in the CGC framework. The parameters in the 
calculation are determined solely by standard nuclear geometry and by 
electron-proton deep inelastic scattering data, which is fit using the 
running coupling BK equation. We show that it is possible to obtain a good 
fit of the HERA inclusive cross section also without an anomalous 
dimension in the initial condition. We argue that one must consistently 
use the proton transverse area as measured by a high virtuality probe in 
DIS also for the single inclusive cross section in proton-proton and 
proton-nucleus collisions. We show that this leads to a midrapidity 
nuclear modification ratio $R_{pA}$ that approaches unity at high 
transverse momentum independently of $\sqrt{s}$, in contrast to most CGC 
calculations in the literature. We also present predictions for future 
forward $R_{pA}$ measurements at the LHC.
\end{abstract}

\maketitle

\section{Introduction}
The Color Glass Condensate (CGC) provides a convenient way to describe strongly
interacting systems in the high energy limit, where nonlinear phenomena,
such as gluon recombination, become important. Because the gluon density scales
as $\sim A^{1/3}$, these nonlinearities are 
enhanced when the target is changed from a proton to a heavy nucleus. 
The p+Pb run at the LHC provides access to a kinematical region 
never explored so far and makes it possible to study QCD phenomena in a 
region where gluon densities are large and non-linear effects become important.

The structure of a hadron can be studied accurately in deep inelastic 
scattering (DIS) where a (virtual) photon scatters off the hadron. 
A large amount of precise high energy electron-proton data measured at HERA
has shown that the gluon density inside a proton grows rapidly at small
Bjorken $x$ or,
equivalently, at high energy. These accurate measurements have also been a 
crucial test for the CGC, and recent analyses have confirmed that the CGC
description is consistent with all the available small-$x$ DIS data~ 
\cite{Albacete:2010sy,Rezaeian:2012ji,Rezaeian:2013tka}.
The CGC makes it possible to consistently describe also other
high-energy hadronic interactions than inclusive deep
inelastic scattering within the same unified framework. These include, for example,
single~\cite{Tribedy:2011aa,Albacete:2012xq,Rezaeian:2012ye} and double inclusive~
\cite{Lappi:2012nh,Albacete:2010pg,Stasto:2012ru,JalilianMarian:2012bd} 
particle production in proton-proton and proton-nucleus collisions,
diffractive DIS~\cite{Kowalski:2006hc,Lappi:2013am} 
and the initial state for the hydrodynamical modeling of a heavy ion 
collision~\cite{Lappi:2011ju,Schenke:2012wb,Gale:2012rq}.
First steps beyond the leading order calculations have also been taken recently
\cite{Balitsky:2008zza,Chirilli:2012jd,Stasto:2013cha}.
Comparing calculations fit to DIS data 
to particle production results for proton-proton and proton-nucleus collisions
at different energies provides a nontrivial test of the
universality of the CGC description, and makes 
predictions for future LHC pA measurements.

In this work we study to what extent it is possible to compute single inclusive 
hadron production in proton-proton and proton-nucleus collisions in a consistent CGC
framework. As an input we use only the HERA data for the inclusive DIS cross section,
which is fitted using the running coupling 
Balitsky-Kovchegov (BK) 
equation~\cite{Balitsky:1995ub,Kovchegov:1999yj,Kovchegov:1999ua}. The resulting 
initial condition for the dipole cross is extended to nuclei using only the standard
Woods-Saxon nuclear density, without any additional nuclear 
parameters. Unlike in some of the recent works using dipole cross sections 
measured in DIS to compute particle production in hadronic collisions, we 
consistently use also the transverse area of the small-$x$ gluonic 
degrees of freedom in a nucleon ($\sigma_0$) determined from the DIS fits. This is
to be contransted with the conceptually separate and numerically much larger
 ``soft'' area of the nucleus given by the total inelastic nucleon-nucleon cross 
section $\sigma_\text{inel}$, which is needed in the Glauber modeling of a nuclear 
collision. Treating the  area factor consistently also 
makes it possible to interpret the ``$K$-factor'' between the normalization 
of the data and our LO calculation as a real indication of the magnitude of 
higher order effects and not only as a completely uncontrolled fit parameter.
 In particular, we will argue that this leads 
to a much  more controlled result for the nuclear modification factor $R_{pA}$ than 
in the existing literature.

This work is structure as follows. In \se \ref{sec:ep} we discuss how one
fits the deep inelastic scattering data in order to obtain an initial condition
for the BK evolution. In \se \ref{sec:sinc} we discuss how 
single inclusive hadron production can be computed in proton-proton 
collisions and generalize the discussion to proton-nucleus case in \se 
\ref{sec:nuke}. Finally we show our numerical results in \se \ref{sec:results}
before concluding in \se \ref{sec:conclusions}.

\section{Electron-proton baseline}
\label{sec:ep}

\begin{figure}[tb]
\begin{center}
\includegraphics[width=0.49\textwidth]{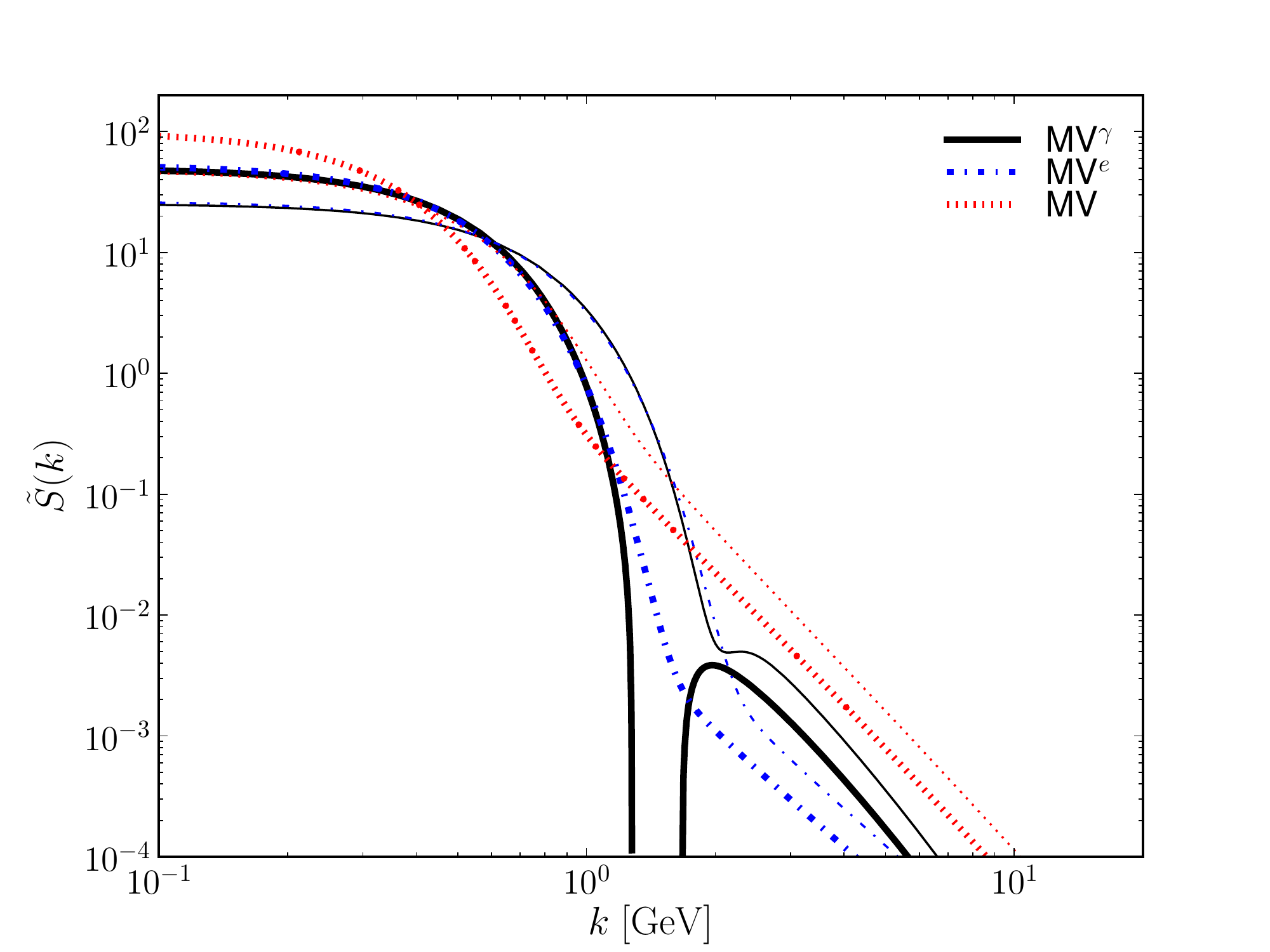}
\end{center}
\caption{
Two dimensional Fourier transform of $S(r)=1-\Ncal(r)$ in fundamental (thick lines) and adjoint (thin lines) representations for MV$^\gamma$, MV$^e$ and MV models.
}\label{fig:ft-s}
\end{figure}

\begin{figure}[tb]
\begin{center}
\includegraphics[width=0.49\textwidth]{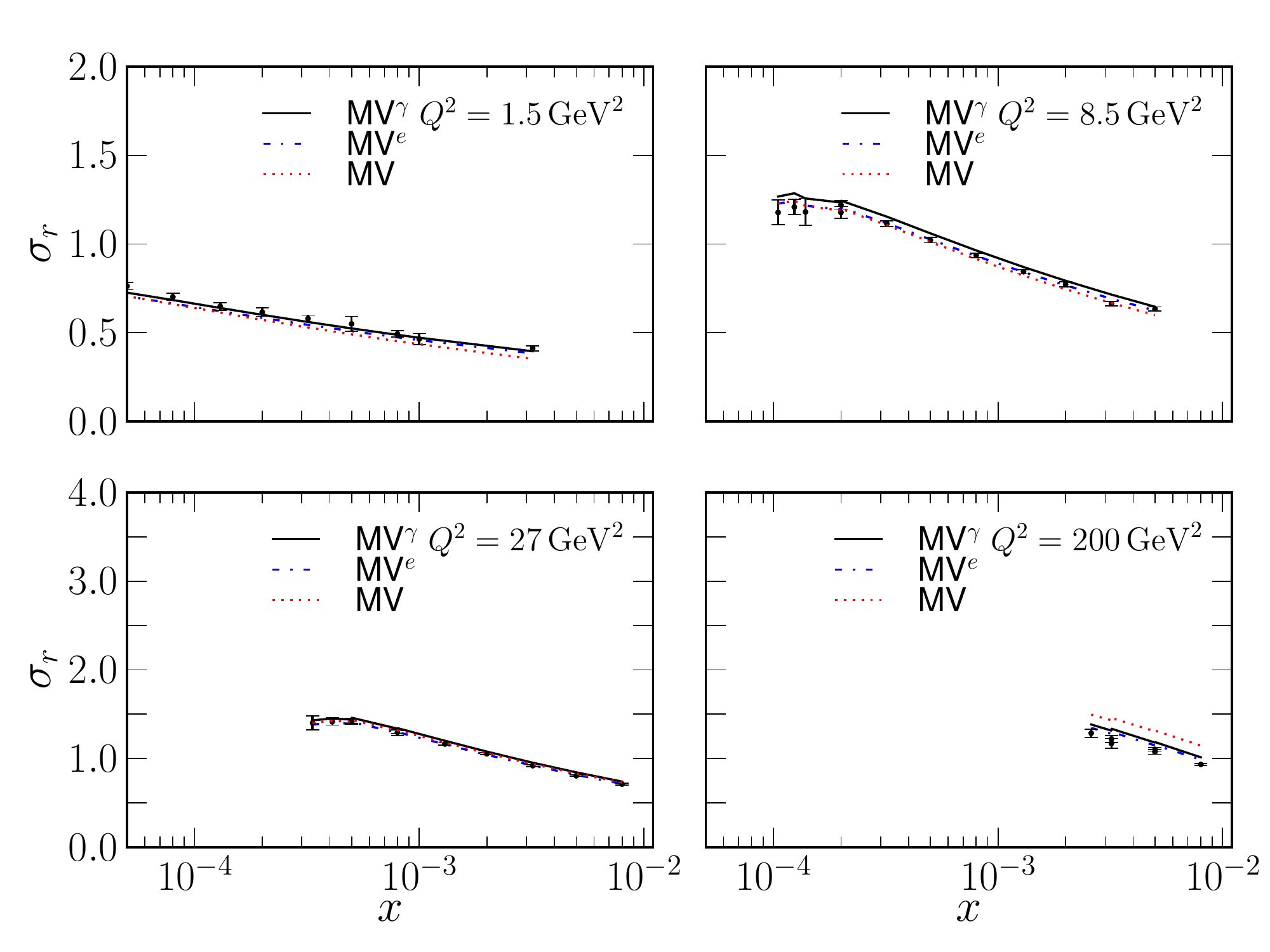}
\end{center}
\caption{
Reduced cross section $\sigma_r$ computed using the MV$^\gamma$, MV$^e$ and MV model initial conditions 
for the dipole amplitude compared with combined HERA 
(H1 and ZEUS) data~\cite{Aaron:2009aa}.\\
}\label{fig:sigmar}
\end{figure}

\begin{figure}[tb]
\begin{center}
\includegraphics[width=0.49\textwidth]{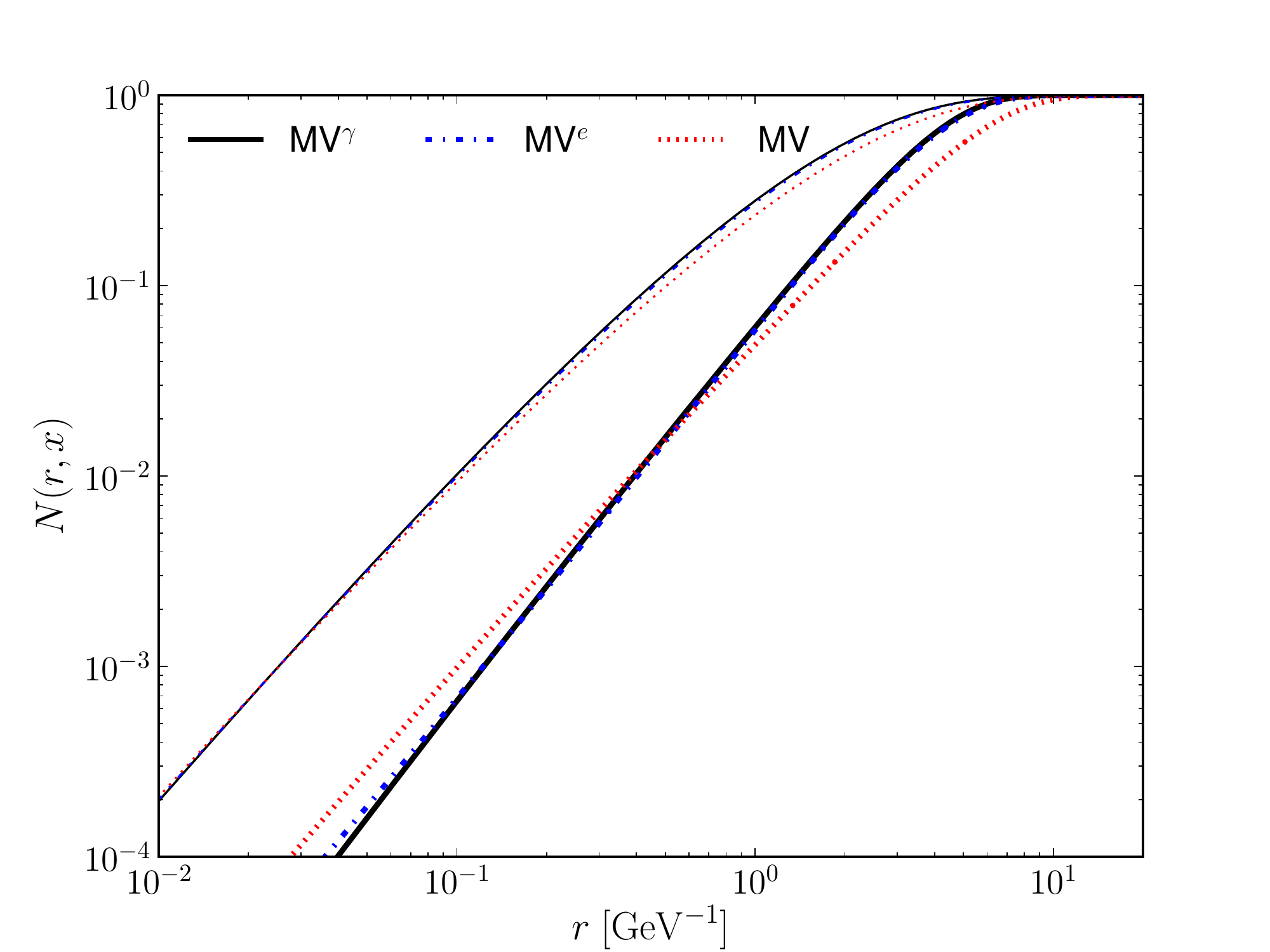}
\end{center}
\caption{
Dipole amplitude at initial $x=10^{-2}$ (thick lines) and after BK evolution at $x=10^{-5}$ (thin lines) from MV$^\gamma$, MV$^e$ and MV models.
}\label{fig:amplitudes}
\end{figure}

Deep inelastic scattering provides a precision measurement of
proton structure. The H1 and ZEUS collaborations have measured the proton
structure functions $F_2$ and $F_L$, and very precise combined results
for the reduced cross section $\sigma_r$ were published 
recently~\cite{Aaron:2009aa}. 

The reduced cross section is a function of the proton structure functions:
\begin{equation}
	\sigma_r(y,x,Q^2) = F_2(x,Q^2) - \frac{y^2}{1+(1-y)^2} F_L(x,Q^2).
\end{equation}
Here $y=Q^2/(sx)$ and $\sqrt{s}$ is the center of mass energy. The structure
functions are related to the virtual photon-proton cross sections 
$\sigma_{T,L}^{\gamma^*p}$ for transverse (T) and longitudinal (L) photons:
\begin{equation}
	F_2(x,Q^2) = \frac{Q^2}{4\pi^2 \aem} (\sigma_T^{\gamma^*p} + \sigma_L^{\gamma^*p})
\end{equation}
and
\begin{equation}
	F_L(x,Q^2) = \frac{Q^2}{4\pi^2 \aem} \sigma_L^{\gamma^* p}.
\end{equation}
In this work we perform a fit to combined HERA $\sigma_r$ 
data~\cite{Aaron:2009aa} for $Q^2<50$ GeV$^2$ and $x<0.01$.

The virtual photon-proton cross 
sections in the Color Class Condensate framework can be computed as
\begin{equation}\label{eq:sigmagammastar}
	\sigma_{T,L}^{\gamma^*p}(x,Q^2) = 2\sum_f \int \der z \int \der^2 \bt |\Psi_{T,L}^{\gamma^* \to f\bar f}|^2 \Ncal(\bt, \rt, x),
\end{equation}
where $\Psi_{T,L}^{\gamma^* \to f\bar f}$ is the photon light cone wave 
function describing how the photon fluctuates to a quark-antiquark pair,
computed from light cone QED~\cite{Kovchegov:2012mbw}. 
The QCD dynamics is 
inside the function $\Ncal(\bt,\rt,x)$ which is the imaginary part of the scattering
amplitude for the process where a dipole (quark-antiquark pair with transverse 
separation $\rt$) scatters off the color field of a hadron
with impact parameter $\bt$. It can not be computed perturbatively, but its
energy (or equivalently Bjorken $x$) dependence satisfies the BK 
equation, for which we use the running
coupling corrections derived in Ref.~\cite{Balitsky:2006wa}. 
The explicit factor 2 in 
\eq\nr{eq:sigmagammastar} comes from the optical theorem, which tells us that 
the total cross section is given by twice the imaginary part of the forward 
scattering amplitude.

We assume here that the impact parameter dependence of the proton 
factorizes and one replaces
\begin{equation}
	2\int \der^2 \bt \to \sigma_0.
\end{equation}
Note that with the usual convention adopted here the factor 2 from the optical theorem 
is absorbed into the constant $\sigma_0$, thus 
the transverse area of the proton is now $\sigma_0/2$.
The proton area $\sigma_0/2$ could in principle  be obtained from diffractive
vector meson production measurements where one can parametrize the transverse
momentum dependence of the production
cross section as $\sim e^{-B_D \Delta_T^2}$, where $\Delta_T$ is the 
momentum transfer in the process.  From the HERA J/$\Psi$ 
data~\cite{Chekanov:2004mw,Aktas:2005xu}
 one obtains $B_D\approx 4$~GeV.
For a gaussian impact parameter profile (exponential in $t$)
these are related by $\sigma_0=4 \pi B_D$.
This would correspond to $\sigma_0\approx 19.5$ mb, which is approximately 
two thirds of the value obtained from the fit to the inclusive data.
For comparison, a theta-function profile, 
in  $b$ with the same $t$-slope at $t=0$ leads to $\sigma_0=8 \pi B_D$, which is 
larger than the value from the inclusive fit.
While the observed $t$-distribution of diffractive 
J/$\Psi$-does not favor a the theta fucntion profile,
it is not known well enough to determine the profile precisely.
We conclude that in practice the measured $t$-distribution
is insufficient to precisely determine $\sigma_0$ alone, and we need to include it 
as a free parameter in the fit.

\begin{table*}
\begin{tabular}{|l||r|r|r|r|r|r|r|}
\hline
Model & $\chi^2/\text{d.o.f}$ &  $\qso^2$ [GeV$^2$] & $\qs^2$ [GeV$^2$] & $\gamma$ & $C^2$ & $e_c$ & $\sigma_0/2$ [mb] \\
\hline\hline
MV & 2.76 & 0.104 & 0.139  & 1 & 14.5 & 1 & 18.81 \\
MV$^\gamma$ & 1.17 & 0.165 & 0.245 & 1.135 & 6.35 & 1 & 16.45 \\
MV$^e$ & 1.15 & 0.060 & 0.238  & 1 & 7.2 & 18.9 & 16.36 \\
\hline
\end{tabular}
\caption{Parameters from fits to HERA reduced cross section data at $x<10^{-2}$ and $Q^2<50\,\mathrm{GeV}^2$ for different initial conditions. Also the corresponding initial saturation scales $\qs^2$ defined via equation $\Ncal(r^2=2/\qs^2)=1-e^{-1/2}$ are shown. The parameters for the MV$^\gamma$ initial condition are obtained by the AAMQS collaboration \cite{Albacete:2010sy}.
}
\label{tab:params}
\end{table*}

As a non-perturbative input one needs also the dipole-proton amplitude at the 
initial $x=x_0$.
For the dipole amplitude we use the following parametrization, based on the
 McLerran-Venugopalan model~\cite{McLerran:1994ni}:
\begin{equation}
\label{eq:aamqs-n}
	\Ncal(\rt) = 1 - \exp \left[ -\frac{(\rt^2 \qso^2)^\gamma}{4} \ln \left(\frac{1}{|\rt| \lqcd}+e_c \cdot e\right)\right],
\end{equation}
where we have generalized the AAMQS~\cite{Albacete:2010sy} form by also
allowing the constant inside the logarithm to be different from $e$. 
This constant plays the role of an infrared cutoff in the MV model
and its value cannot be fixed by a weak coupling calculation, it is therefore
natural to leave it as a free parameter. The other fit parameters are
anomalous dimension $\gamma$ and initial saturation scale $\qso^2$.

The BK equation with running coupling requires the strong coupling constant 
$\as$ as a function of the transverse separation $r=|\rt|$. In order to obtain 
a slow enough evolution to be compatible with the data (see discussion in 
Ref.~\cite{Kuokkanen:2011je}) we include, as in \cite{Albacete:2010sy}, 
an additional fit parameter $C^2$ such that
\begin{equation}
	\as(r) = \frac{12\pi}{(33 - 2N_f) \log \left(\frac{4C^2}{r^2\lqcd^2} \right)},
\end{equation}
with $\lqcd$ fixed to the value $0.241$ GeV.
It has been argued that this encodes the uncertainty for the scale
at which the coordinate space strong coupling constant should be evaluated.
On the other hand it has been justified analytically~\cite{Kovchegov:2006vj} and
confirmed numerically~\cite{Lappi:2012vw} (for  a slightly different running
coupling prescription and for the JIMWLK equation) that performing
the Fourier-transform would lead to
$C^2=e^{-2\gamma_e}$. With this interpretation 
the fit result $C^2 \sim 6$ corresponds to the QCD scale taking the value
  $\lqcd/C \sim 50$~MeV.
In numerical solutions we freeze $\as$ to $0.7$ in the infrared. The quark mass
is fixed to $m=0.14$ GeV, as it was found in Ref. \cite{Albacete:2010sy} that
taking it to be a fit parameter does not improve the fit quality significantly.
We also consider only the three light quarks in this work.

The unknown parameters are obtained by performing a fit to small-$x$ DIS
data. The first parametrization considered in this work, denoted by 
MV$^\gamma$, is obtained by setting $e_c \equiv 1$ in \eq \eqref{eq:aamqs-n}
but keeping the anomalous dimension $\gamma$ as a fit parameter.
A global fit to HERA, NMC and E665 deep inelastic scattering data for this
parametrization was performed by the AAMQS collaboration in 
Ref.~\cite{Albacete:2010sy}, resulting in a very good fit 
$\chi^2/\text{d.o.f}\approx 1.17$. The fit parameters are listed in Table 
\ref{tab:params}.

We fit two other initial conditions for the dipole amplitude to the HERA 
reduced cross section data. First, we consider a parametrization where we
do not include anomalous dimension ($\gamma \equiv 1$) but let $e_c$ to be a 
free fit parameter. The second model studied for comparison is the MV model
without modifications, where $\gamma \equiv 1$ and $e_c \equiv 1$. As we do not
include E665 or NMC data, the fit procedure is not exactly the same as in Ref.~
\cite{Albacete:2010sy}. However, due to its very small errors the HERA data 
dominates the AAMQS fit, and this 
results in only very minor differences between our result and that of 
Ref.~\cite{Albacete:2010sy}\footnote{The other difference between our approach 
and Ref.~\cite{Albacete:2010sy} is that for simplicity we do use the redefinition
of $x\to x(1+4m_f^2/Q^2)$, where $m_f$ is the quark mass.}. Our fit result for 
the MV$^\gamma$ model is 
$\qso^2=0.159$ GeV$^2, \gamma=1.129, C^2=7.05$ and $\sigma_0/2=16.35$ fm.
We have also confirmed numerically that the two different MV$^\gamma$ model
parametrizations give basically the same result when computing quantities
considered in this work.

Our second parametrization, denoted by MV$^{e}$ here, 
has an infrared cutoff $e_c$ as a fit parameter but no anomalous dimension 
($\gamma \equiv 1$). The fit quality is essentially as good as for the
MV$^\gamma$ model, with the best fit giving $\chi^2/\text{d.o.f}\approx 1.15$.
The fit parameters are listed in Table \ref{tab:params}.
Viewed in momentum space
this parametrization provides a smoother interpolation 
between the small-$k$ saturation region (where it resembles the Gaussian
GBW form) and the power law at behavior high $k$. This is demonstrated
in Fig. \ref{fig:ft-s}, where we show the Fourier-trasform of 
$S(\rt) = 1-\Ncal(\rt)$, which is proportional to the ``dipole'' gluon distribution.


For comparison we also study a pure MV model initial condition fixing 
$\gamma=1$ and $e_c=1$, resulting in the  parameters
listed in Table \ref{tab:params}. 
The fit quality is not as good as with the modified MV model, 
($\chi^2/\text{d.o.f}\sim 2.8$), but as one can see from Fig. \ref{fig:sigmar}, the 
description of the small-$x$ DIS data is still reasonable.

The dipole amplitudes at initial Bjorken-$x$ obtained from the MV$^\gamma$ and 
MV$^e$ models are close to each other, and they both deviate significantly 
from the pure MV model. To demonstrate this, we show in Fig. 
\ref{fig:amplitudes} the dipole amplitudes $\Ncal(\rt, x)$ at initial $x=x_0=0.01$
and at $x=10^{-5}$.
Note that one can not directly compare
the values of the parameters $\qso^2$ in the initial condition, as the functional
form in different parametrizations is different.
To perform a model-independent comparison of the values of the saturation 
scale in different fits, we define
the saturation scale $\qs^2$ as a solution to the equation 
$\Ncal(\rt^2=2/\qs^2)=1-e^{-1/2}$. The initial saturation scales with this 
model-independent definition 
are also summarized in Table \ref{tab:params}.

\section{Single inclusive hadron production in CGC}
\label{sec:sinc}
The spectrum of gluons can be computed using 
$k_T$-factorization~\cite{Kovchegov:2001sc} which, for $k_T\gtrsim \qs$
 is known to give the same
gluon spectrum than what is obtained by solving the classical Yang-Mills 
equations~\cite{Blaizot:2010kh}:
\begin{multline}
\label{eq:ktfact-bdep}
\frac{\der \sigma}{\der y \der^2 \kt \der^2 \bt} = \frac{2 \as}{\cf \kt^2 } \int \der^2 \qt \der^2 \st \frac{\varphi_p(\qt,\st)}{\qt^2} 
	\\ \times \frac{\varphi_p(\kt-\qt,\bt-\st)}{(\kt-\qt)^2}.
\end{multline}
Here $\varphi_p$ is the dipole unintegrated gluon
distribution (UGD) of the 
proton~\cite{Kharzeev:2003wz,Blaizot:2004wu,Dominguez:2011wm} 
and $\bt$ is the 
impact parameter. For the proton we assume that the impact parameter dependence 
factorizes and 
\begin{equation}
\label{eq:ugd}
	\varphi_p(\kt) = \int \der^2 \bt \varphi_p(\kt, \bt) 
	 = \frac{\cf \sigma_0/2}{8\pi^3 \as} \kt^4 \tilde S^p(\kt).
\end{equation}
Here $\tilde S^p(k)$ is the two dimensional Fourier transform of the 
dipole-proton scattering matrix $S^p(r)=1-\Ncal^p_A(r)$, where $\Ncal^p_A$ is the
dipole-proton scattering amplitude in adjoint representation: 
$\Ncal_A = 2\Ncal-\Ncal^2$.
For the proton DIS area $\sigma_0/2$ we use the value from the fits to DIS data, 
see \se \ref{sec:ep}. 

Let us now consider a proton-proton collision. The cross section is obtained by 
integrating \eq \eqref{eq:ktfact-bdep} over the impact parameter, which gives
\begin{multline}
	\frac{\der \sigma}{\der y \der^2 \kt} = \frac{(\sigma_0/2)^2}{(2\pi)^2} \frac{\cf}{2\pi^2 \kt^2 \as} \int \frac{\der^2 \qt}{(2\pi)^2} \qt^2 \tilde S^p(\qt)
	\\ \times  (\kt-\qt)^2 \tilde S^p(\kt-\qt).
\end{multline}
The invariant yield is defined as the production cross section 
divided by the total inelastic cross section $\sigma_\text{inel}$ and  thus becomes
\begin{multline}
\label{eq:ktfact-pp}
	\frac{\der N}{\der y \der^2 \kt} = \frac{(\sigma_0/2)^2}{\sigma_\text{inel}} \frac{\cf}{8\pi^4  \kt^2 \as} \int \frac{\der^2 \qt}{(2\pi)^2} \qt^2 \tilde S^p(\qt) \\
		\times (\kt-\qt)^2 \tilde S^p(\kt-\qt).
\end{multline}

\begin{figure}[tb]
\begin{center}
\includegraphics[width=0.49\textwidth]{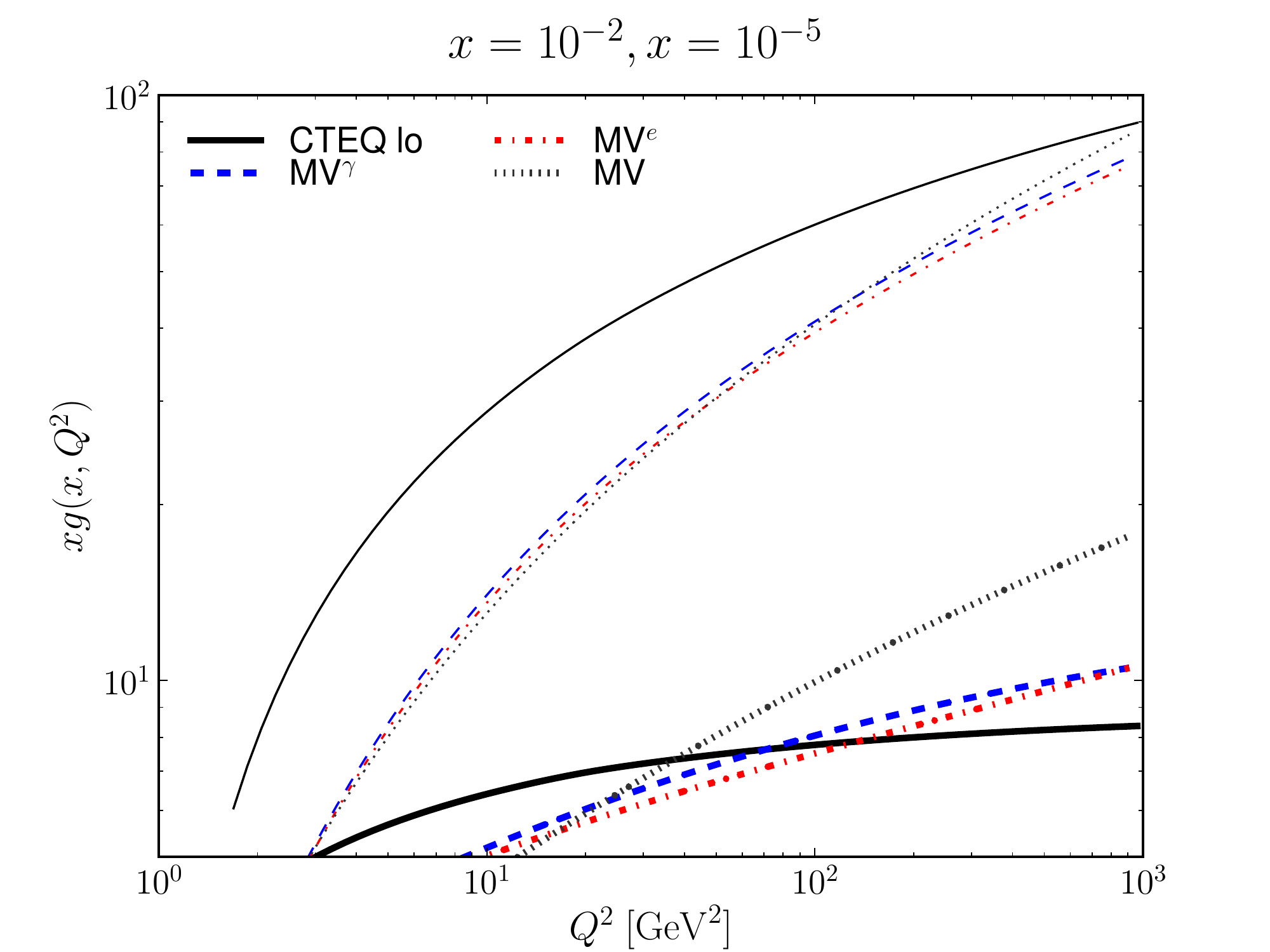}
\end{center}
\caption{
Gluon distribution function at $x=10^{-2}$ (lower thick lines) and at $x=10^{-4}$ (upper thin lines) computed using MV, MV$^\gamma$ and MV$^e$ initial conditions compared with the leading order CTEQ gluon distribution. }
\label{fig:xg_x}
\end{figure}

Assuming that $|\kt|$ is much larger than the saturation scale of one of the
protons we obtain the hybrid formalism result
\begin{equation}
	\label{eq:pp-hybrid}
	\frac{\der N}{\der y \der^2 \kt} = \frac{\sigma_0/2}{\sigma_\text{inel}} \frac{1}{(2\pi)^2} xg(x,\kt^2) \tilde S^p(\kt),
\end{equation}
where
\begin{equation}
	\label{eq:xg}
	xg(x,\kt^2) = \int_0^{\kt^2} \frac{\der \qt^2}{\qt^2} \varphi_p(\qt)
\end{equation}
is the integrated gluon distribution function. This can then 
be replaced by the conventional parton distribution function, for which we can use the 
CTEQ LO~\cite{Pumplin:2002vw} pdf. In Fig. \ref{fig:xg_x} we show
this function as resulting from the dipole fits to HERA data 
at $x=10^{-2}$ and $x=10^{-4}$ 
compared  with the leading order CTEQ distribution.
At the initial $x=10^{-2}$ the gluon density grows much faster as a function
of $Q^2$ when the MV model is used, and the results obtained using the
 MV$^\gamma$ and MV$^e$ models are close to each other. 
This observation suggests a new interpretation for why the experimental 
data seems to favor a steeper initial condition for BK evolution
such as  MV$^\gamma$ and MV$^e$: this form provides a better parametrization 
of the large logarithms of $Q^2$ that are resummed by DGLAP evolution, 
but not included in  the leading order BK equation.
When moving
from the initial condition to smaller $x$ the gluon density from the CTEQ 
distribution grows faster than what is obtained from the BK evolution.
At smaller $x$ the difference between the initial conditions is relatively 
small as the solutions of the BK equation approach the universal form.

HERA measurements of diffractive vector meson 
electroproduction~\cite{Aaron:2009xp} indicate that the proton transverse area 
measured with a high virtuality probe is smaller than in soft interactions. 
In our case this shows up
as a large difference in the numerical values of 
$\sigma_0/2$ and $\sigma_\text{inel}$, and leads to an energy 
dependent factor
$\frac{\sigma_0/2}{\sigma_\text{inel}}\sim 0.2\dots 0.3$ in the 
particle yield~\nr{eq:pp-hybrid}, in contrast with the treatment  
often used in CGC calculations.
Physically this corresponds  to a two-component picture
of the transverse structure of the nucleon (see also
Ref.~\cite{Frankfurt:2010ea} for a very similar discussion).
The small-$x$ gluons responsible for semihard particle production
occupy a small area $\sim \sigma_0/2$ in the core of the nucleon. 
This core is surrounded by a nonperturbative edge that becomes larger
with $\sqrt{s}$, but only participates in soft interactions that contribute
to the large total inelastic cross section $\sigma_\text{inel}$
(we use $\sigma_\text{inel}=42$ mb at RHIC~\cite{Adams:2006uz}, 
$\sigma_\text{inel}=44.4$ mb at Tevatron~\cite{Aaltonen:2009ne}
and $\sigma_\text{inel}=70$ mb at LHC energies~\cite{Antchev:2011vs}).
This description of the transverse profile in terms of only two numbers, 
an energy-independent $\sigma_0$ and an energy-dependent $\sigma_\text{inel}$
is of course very simplistic, but we believe it captures a physical 
feature that has been neglected in many works on the subject.
Note that Ref.~\cite{Tribedy:2010ab} models the same physics 
by consistently using a $b$-dependent dipole cross section, 
and incorporating the soft physics as a $\sqrt{s}$-dependent
upper limit in the integration over $b$.
We will in Sec.~\ref{sec:nuke} show that this separation between 
the two transverse areas brings much clarity to the extension of 
the calculation from protons to nuclei.

Now that also the normalization ($\sigma_0/2$) from HERA data is used in the 
calculation of the single inclusive spectrum, the result represents the  actual LO CGC 
prediction for also the normalization of the spectrum. As is often
the case for perturbative QCD, the LO result only agrees with data within a factor of 
$\sim 2$.  
We therefore  multiply the resulting spectrum with a 
``$K$-factor'' to bring it to the level of the experimental data. 
Now that the different areas $\sigma_0$ and $\sigma_\text{inel}$ are properly
included, this factor has a more conventional interpretation of 
the expected effect of NLO corrections on the result; although it depends
quite strongly on the fragmentation function. In particular we note that
the numerical values cannot be directly compared with the numerical values
given in Ref.~\cite{Albacete:2012xq}.


In order to obtain a hadron spectrum from the parton spectrum we calculate 
convolution with the DSS LO fragmentation function~\cite{deFlorian:2007aj} and,
when using the hybrid formalism, also add the light quark-initiated channel
to the gluonic one in \eq\nr{eq:pp-hybrid}. The momentum scale 
for the parton distribution functions, fragmentation functions and the 
strong coupling constant $\as$ are chosen as the transverse 
momentum of the produced hadron.

\section{From proton to nucleus}
\label{sec:nuke}

Due to a lack of small-$x$ nuclear DIS data we can not perform a
 similar fit to nuclear targets than what is done with the proton. 
Instead we use the optical Glauber model to generalize our dipole-proton 
amplitude to dipole-nucleus scattering. 

First we observe that the total dipole (size $r$)-proton 
cross section reads
\begin{equation}
	\sigma_\text{dip}^p = \sigma_0 \Ncal^p(r).
\end{equation}
In the dilute limit of very small dipoles the dipole-nucleus cross section 
should be just an incoherent sum of dipole-nucleon cross sections, i.e.
$\sigma_\text{dip}^A = A\sigma_\text{dip}^p$.
 On the other hand for large 
dipoles we should have $\der \sigma_\text{dip}^A / \der^2 \bt 
\equiv 2 \Ncal^A(\rt,\bt) \leq 2$. These 
requirements are satisfied with an exponentiated dipole-nucleus scattering amplitude
\begin{equation}\label{eq:glaubersigmadipa}
\Ncal^A(\rt,\bt) = \left[ 1 - \exp\left( -\frac{AT_A(\bt)}{2} \sigma_\text{dip}^p \right) \right].
\end{equation}
This form is an average of the dipole cross section over the fluctuating
positions of the nucleons in the nucleus (see e.g. \cite{Kowalski:2007rw}), 
and thus incorporates in an analytical expression the fluctuations discussed 
e.g. in Ref.~\cite{Albacete:2012xq}.

Using the form \nr{eq:glaubersigmadipa} directly in computing particle production is, 
however problematic. Because the forward $S$-matrix element
 $S=1-\Ncal^A(\rt,\bt)$ approaches
a limiting value $\exp\left( -\frac{AT_A(\bt)}{2} \sigma_0 \right) 
\sim \exp\left(-A^{1/3}\right) $ and not exactly zero at large $r$, the dipole 
gluon distribution  develops unphysical oscillations as a function of $k$.
We therefore expand the proton-diple cross section in  \eq\nr{eq:glaubersigmadipa} 
and use the approximation 
\begin{equation}
	\sigma_\text{dip}^p = \sigma_0 \Ncal^p(\rt) \approx \sigma_0 \frac{(\rt^2 \qso^2)^\gamma}{4} \ln \left(\frac{1}{|\rt|\lqcd}+e_c \cdot e\right)
\end{equation}
in the exponent of \eq\nr{eq:glaubersigmadipa}.
The dipole-nucleus amplitude is then obtained by solving the rcBK evolution 
equation with an initial condition
\begin{multline}\label{eq:ainitc}
	\Ncal^A(\rt,\bt) = 1 - \exp\left[ -A T_A(\bt) \frac{\sigma_0}{2} \frac{(\rt^2 \qso^2)^\gamma}{4} \right.  \\
	\left. \times \ln \left(\frac{1}{|\rt|\lqcd}+e_c \cdot e\right) \right].
\end{multline}
We emphasize that besides the Woods-Saxon nuclear density $T_A(\bt)$, all the
parameters in this expression result from the fit to HERA data. Among recent
works on the subject this can be 
contrasted with e.g. Ref.~\cite{Albacete:2012xq} where the area corresponding to 
$\sigma_0/2$ in \eq\nr{eq:ainitc} is set by fiat to $42$~mb, or to 
Ref.~\cite{Rezaeian:2012ji}, where initial saturation scale is varied within
a large range. The ``optical Glauber'' initial condition \nr{eq:ainitc} also
brings to evidence the advantage of the MV$^e$ parametrization, which  
achieves a good fit to HERA data while imposing $\gamma=1$.
 In contrast to the MV$^\gamma$ fit, this functional form avoids the ambiguity
encountered in e.g.~\cite{Albacete:2012xq}
of whether the factor $AT_A(\bt) \sigma_0/2$ should be replaced by 
$(AT_A(\bt) \sigma_0/2)^\gamma$ to achieve a natural scaling of 
$\qs^2$  with the nuclear thickness.

\begin{figure}[tb]
\begin{center}
\includegraphics[width=0.49\textwidth]{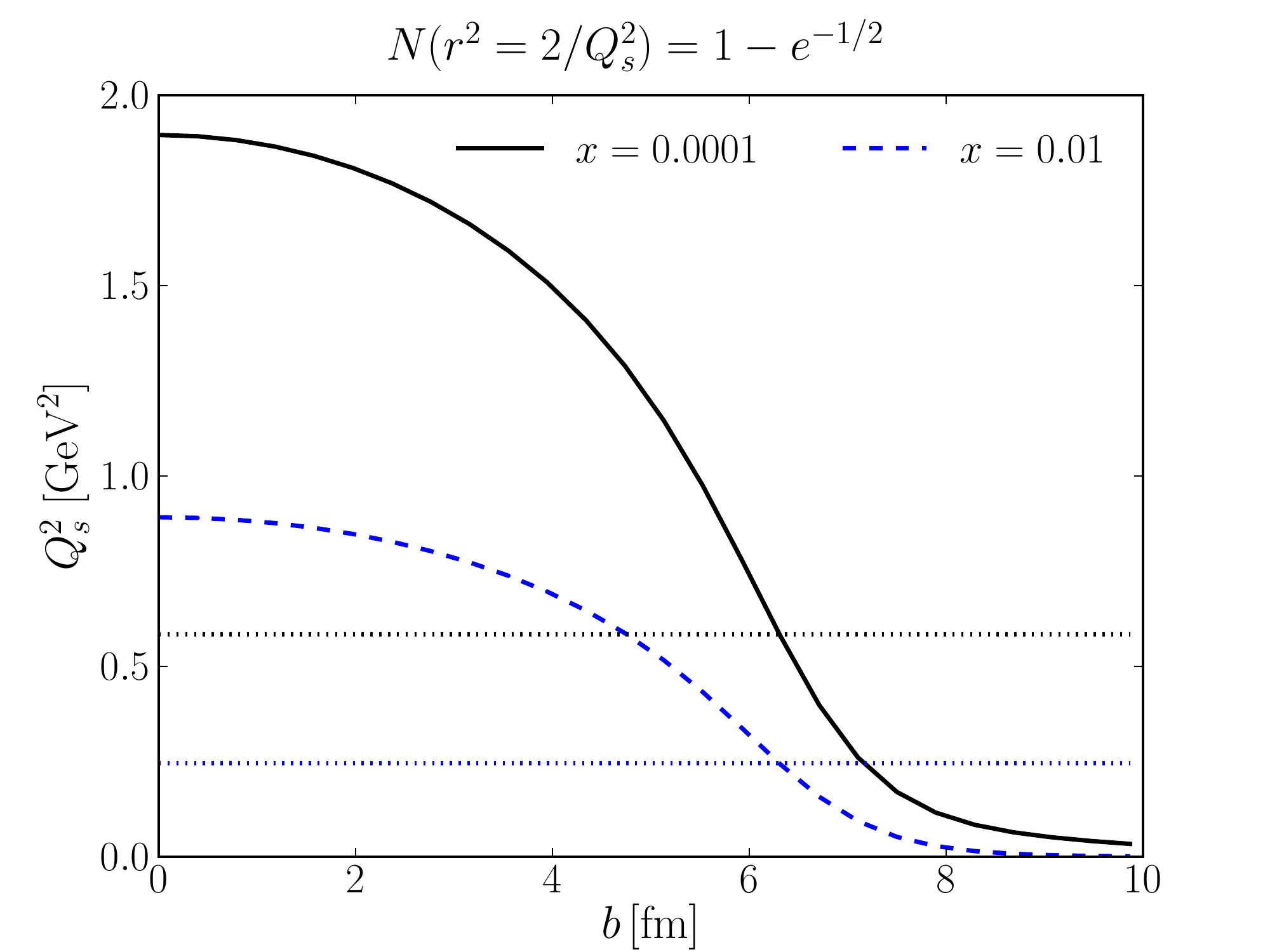}
\end{center}
\caption{
Saturation scale of the lead nucleus at $x=10^{-2}$ and at $x=10^{-4}$ as a function of impact parameter obtained using the MV$^\gamma$ initial condition. The dashed lines show the saturation scale of the proton at the same values of $x$. 
}\label{fig:qs_b}
\end{figure}

The fully impact parameter dependent BK equation develops unphysical Coulomb
 tails which would need an additional screening mechanism at the confinement scale
(see e.g.~\cite{GolecBiernat:2003ym,Berger:2010sh,Berger:2011ew,Berger:2012wx}). We 
therefore  solve the scattering amplitudes for each $\bt$ independently. 
Due to the rapid increase of the scattering amplitude at low densities 
(large $|\bt|$) this effectively causes the nucleus to grow rapidly on the 
edges at large energies. To demonstrate this we plot in Fig. \ref{fig:qs_b}
 the saturation scale $\qs^2$ of the nucleus as a function of impact parameter
 $\bt$, using again the model-independent definition of $\qs$ as the solution of
$\Ncal(\rt^2=2/\qs^2, \bt)=1-e^{-1/2}$. The saturation scale of
 the lead nucleus falls below the proton saturation scale at 
$|\bt| \gtrsim 6.3\,\mathrm{fm}$, which corresponds to centrality
 $\gtrsim 70\%$, see Table \ref{tab:centrality_lhc}. Due to the
 unphysical increase of the gluon density at very large $|\bt|$ 
we do not consider this parametrization to be reliable in that region. Instead,
for calculating minimum bias observables we simply scale up from 
proton-proton collisions by assuming that $R_{pA}=1$ for very large impact 
parameters (see below).
A more refined treatment of the nuclear edge would be possible by replacing 
the optical Glauber framework by a Monte Carlo one among the lines of 
\cite{Albacete:2012xq}, but the edge region gives a very small contribution 
to total inclusive cross sections and this would have a negligible effect on
 the observables considered in this paper.

When computing proton-nucleus cross sections we compute convolutions of the 
nuclear and the proton unintegrated gluon distributions. In terms of the 
dipole amplitudes the $k_T$ factorization formula now reads
\begin{multline}
	\frac{\der N(\bt)}{\der y \der^2 \kt} = \frac{\sigma_0/2}{(2\pi)^2} \frac{\cf}{2\pi^2 \kt^2 \as} \int \frac{\der^2 \qt }{(2\pi)^2} \qt^2 \tilde S^p(\qt) \\
	\times (\kt-\qt)^2 \tilde S^A(\kt-\qt),
\end{multline}
where $\tilde S^p$ and $\tilde S^A$ are Fourier transforms of the dipole-proton 
and dipole-nucleus scattering matrices, respectively. Assuming moreover that 
the transverse momentum of the produced parton is much larger than the proton 
saturation scale we get the hybrid formalism result
\begin{equation}
	\frac{\der N(\bt)}{\der y \der^2 \kt} = \frac{1}{(2\pi)^2} xg(x,\kt^2) \tilde S^A(\kt).
\end{equation}
Notice that, in contrast to \eq\nr{eq:pp-hybrid} in this case we do not get a factor
$(\sigma_0/2)/\sigma_\text{inel}$ in the yield.

Let us then show that with this parametrization we get $R_{pA}\to 1$ at large
transverse momenta. First we observe that at large $|\kt|$ (when also $x$ approaches
the initial condition value) the particle yield in
proton-nucleus collisions is
\begin{multline}
\label{eq:pa-yield}
\der N^{pA} \sim xg \tilde S^A(\kt) \\ \sim xg \int \der^2 \rt e^{i\kt \cdot \rt} \exp\left( -\frac{AT_A(b)}{2} \sigma_\text{dip}^p\right)\\
	\sim xg A T_A(b) \frac{\sigma_0}{2} \Ncal^p.
\end{multline}
On the other hand in proton-proton collisions we get
\begin{equation}
	\der N^{pp} \sim \frac{\sigma_0/2}{\sigma_\text{inel}} xg \Ncal^p.
\end{equation}
Now as $N_\text{bin}=AT_A \sigma_\text{inel}$, the nuclear modification ratio is
\begin{equation}
	R_{pA} = \frac{\der N^{pA}}{N_\text{bin}\der N^{pp}} \to 1
\end{equation}
at all $\sqrt{s}$, even as $\sigma_\text{inel}$ and thus $N_\text{bin}$
are changing with  $\sqrt{s}$ while the initial saturation scale $\qso$ is not.
This is in marked contrast to e.g. 
Refs.~\cite{Albacete:2010bs,Tribedy:2011aa,Albacete:2012xq}), where the physics of
high energy evolution is basically the same, but the treatment of the transverse
geometry different, resulting in a variety of very different predictions
for the high transverse momentum behavior of $R_{pA}$.


Once the dipole-nucleus amplitude is known, one immediately gets the 
unintegrated gluon distribution  of the nucleus at fixed impact paramter $\bt$
from \eq \eqref{eq:ugd}:
\begin{equation}
	\varphi^A(\kt, \bt) = \frac{\cf}{8\pi^3 \as} \kt^4 \tilde S^A(\kt,\bt).
\end{equation} 

In order to obtain estimates for the dipole-nucleus scattering amplitude at 
large impact parameters we assume that in the region where the saturation 
scale  of the nucleus would fall below the corresponding scale of the proton,
at $|\bt|>b_0$, we obtain the differential particle production yield in 
proton-nucleus collision by using the expanded form \eq\nr{eq:pa-yield}
at all $r$. This expansion 
is justified because at large impact parameters the combination  $AT_A(b)$ is small.
This gives, substituting $AT_A = N_\text{bin}/\sigma_\text{inel}$,
\begin{equation}
	\der N^{pA}  \sim xg N_\text{bin} \frac{\sigma_0/2}{\sigma_\text{inel}} N^p 
	= N_\text{bin} \der N^{pp}.
\end{equation}
Notice that this parametrization is equivalent to imposing
 $R_{pA}=1$ at large impact parameters.

In proton-nucleus collisions it is not possible to determine the impact 
parameter by measuring the total multiplicity as well as in heavy ion
collisions, due to the large multiplicity fluctuations for a fixed impact parameter.
 The first LHC proton-lead results are
divided to centrality classes based on multiplicity or energy deposit in 
forward calorimeters. This is theoretically difficult quantity to handle,
so we assume that we can obtain reasonable estimates for different centrality 
classes by using a standard optical Glauber model described briefly in 
Appendix \ref{glauber}.

\section{Results}
\label{sec:results}

\begin{figure}[tb]
\begin{center}
\includegraphics[width=0.49\textwidth]{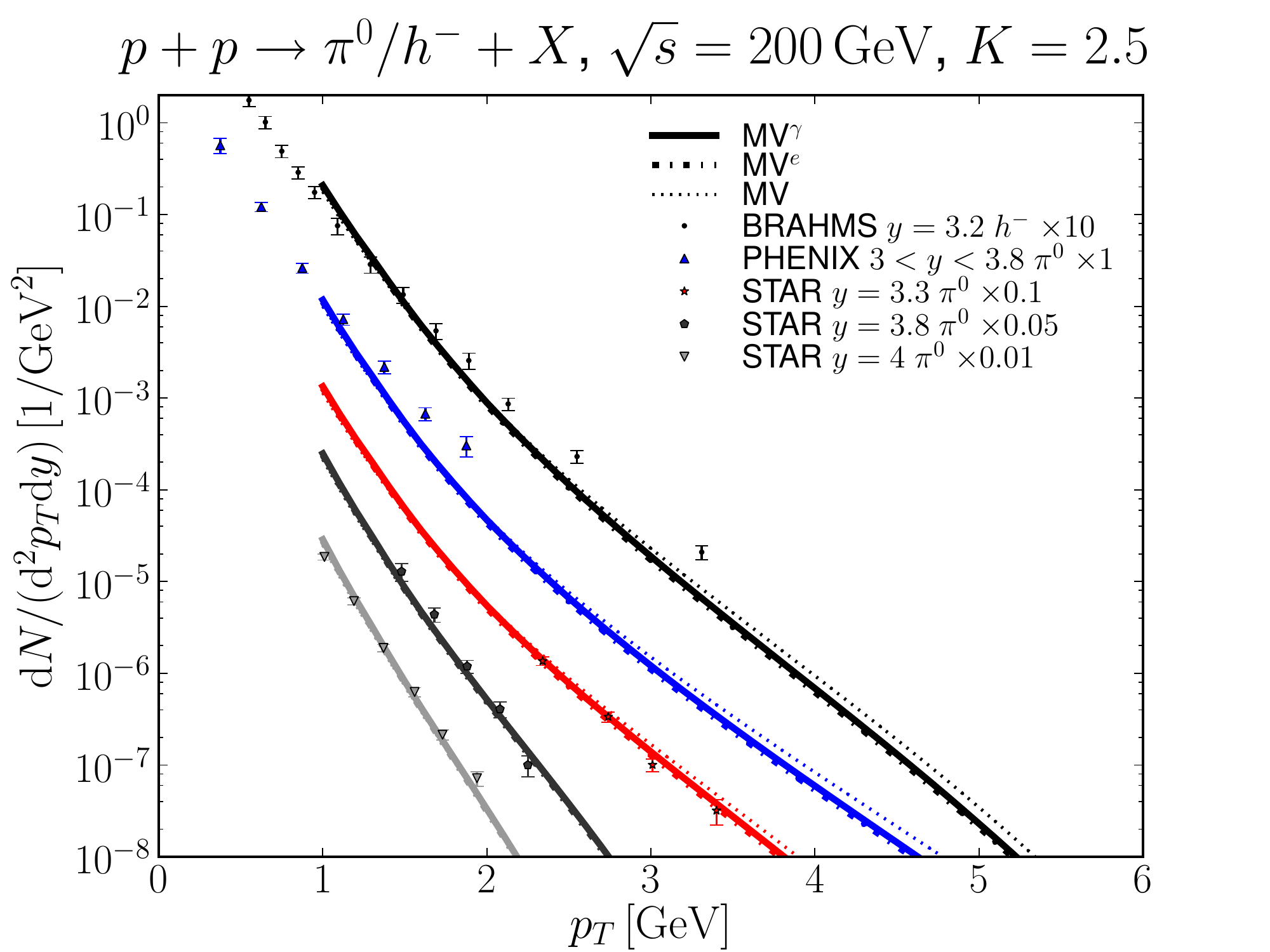}
\end{center}
\caption{
Single inclusive $\pi^0$ and negative hadron production computed using MV, MV$^e$ and MV$^\gamma$ initial conditions compared with RHIC data from STAR~\cite{Adams:2006uz}, PHENIX~\cite{Adare:2011sc} and BRAHMS~\cite{Arsene:2004ux} collaborations. \\
}\label{fig:rhic_pp_yield}
\end{figure}

\begin{figure}[tb]
\begin{center}
\includegraphics[width=0.49\textwidth]{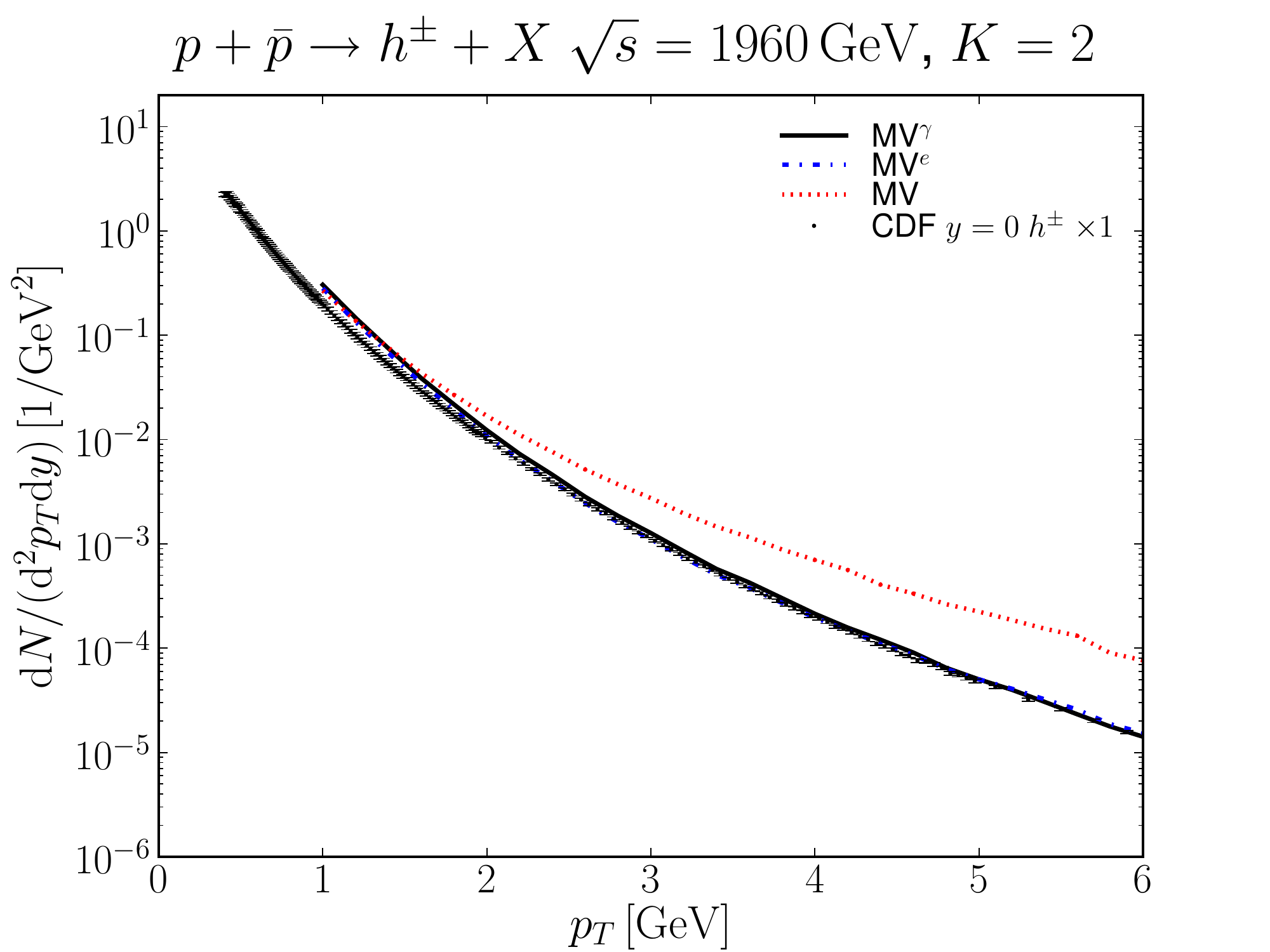}
\end{center}
\caption{
Single inclusive charged hadron production computed using MV, MV$^e$ and MV$^\gamma$ initial conditions compared with CDF data~\cite{Aaltonen:2009ne}.
}\label{fig:tevatron_yield}
\end{figure}

\begin{figure}[tb]
\begin{center}
\includegraphics[width=0.49\textwidth]{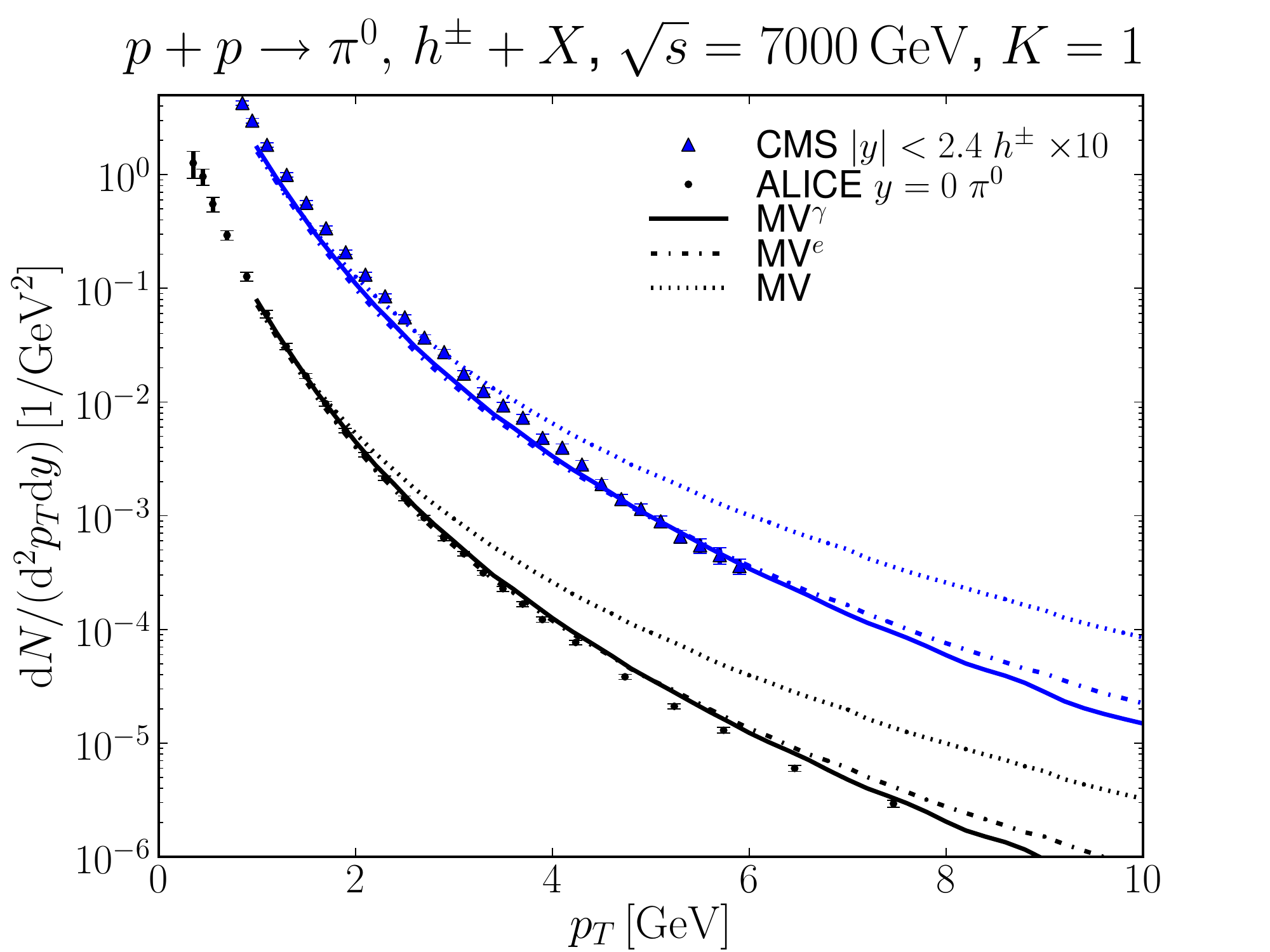}
\end{center}
\caption{
Single inclusive $\pi^0$ production computed using MV, MV$^\gamma$ and MV$^e$ initial conditions at $\sqrt{s}=7000$ GeV compared with ALICE $\pi^0$ \cite{Abelev:2012cn} and CMS charged hadron data~\cite{Khachatryan:2010us}. The CMS yield is computed at $y=0$.\\
}\label{fig:alice_cms_pp_yield}
\end{figure}

\begin{figure}[tb]
\begin{center}
\includegraphics[width=0.49\textwidth]{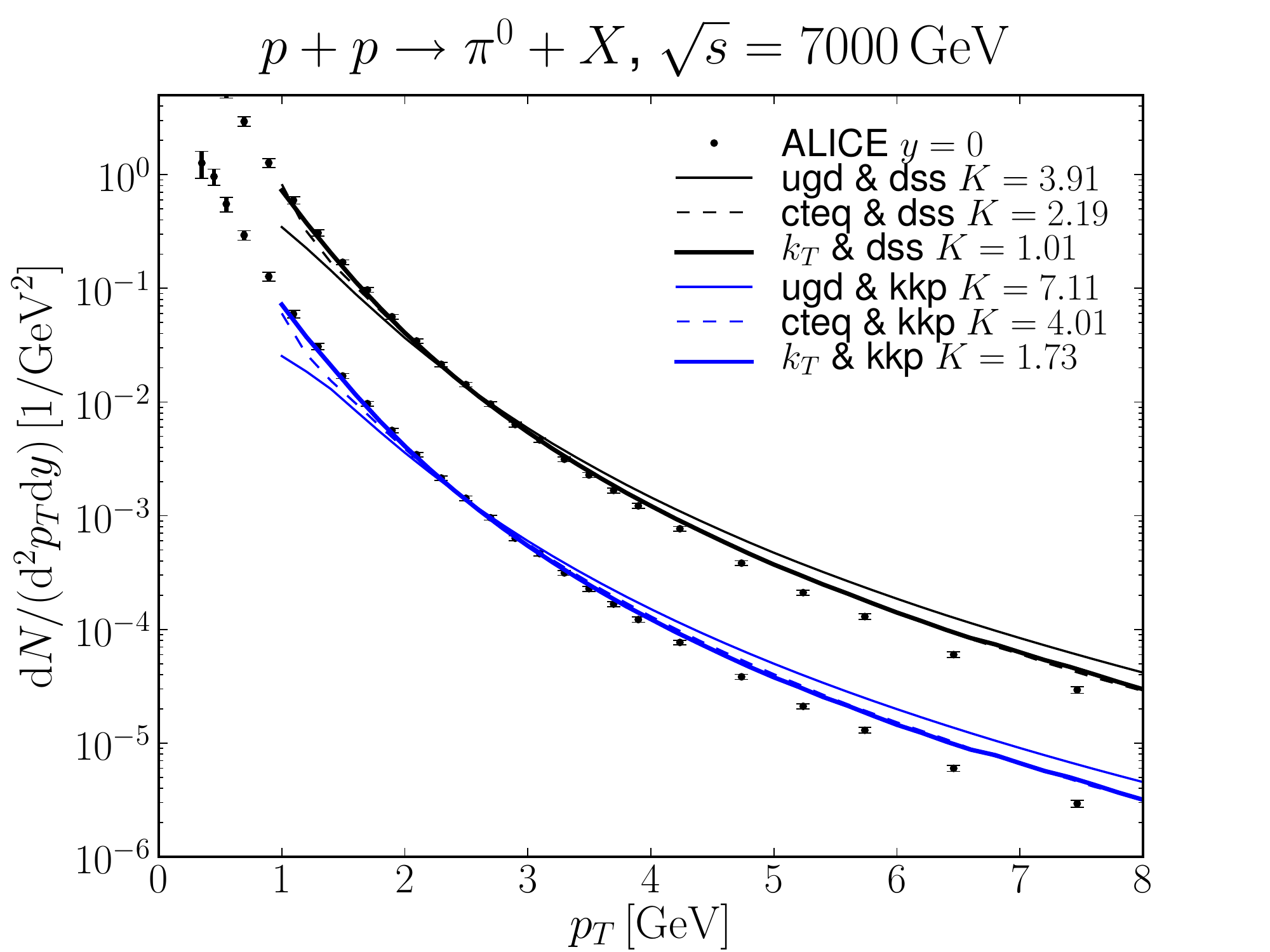}
\end{center}
\caption{
Single inclusive $\pi^0$ production at $\sqrt{s}=7000$ GeV compared with ALICE $\pi^0$ \cite{Abelev:2012cn}  data computed using $k_T$-factorization and the hybrid formalism with CTEQ and UGD parton distribution functions and DSS (upper curves, multiplied by $10$) and KKP (lower curves) fragmentation functions. The initial condition for the BK evolution is MV$^e$.}\label{fig:alice_pi0_models}
\end{figure}

\begin{figure}[tb]
\begin{center}
\includegraphics[width=0.49\textwidth]{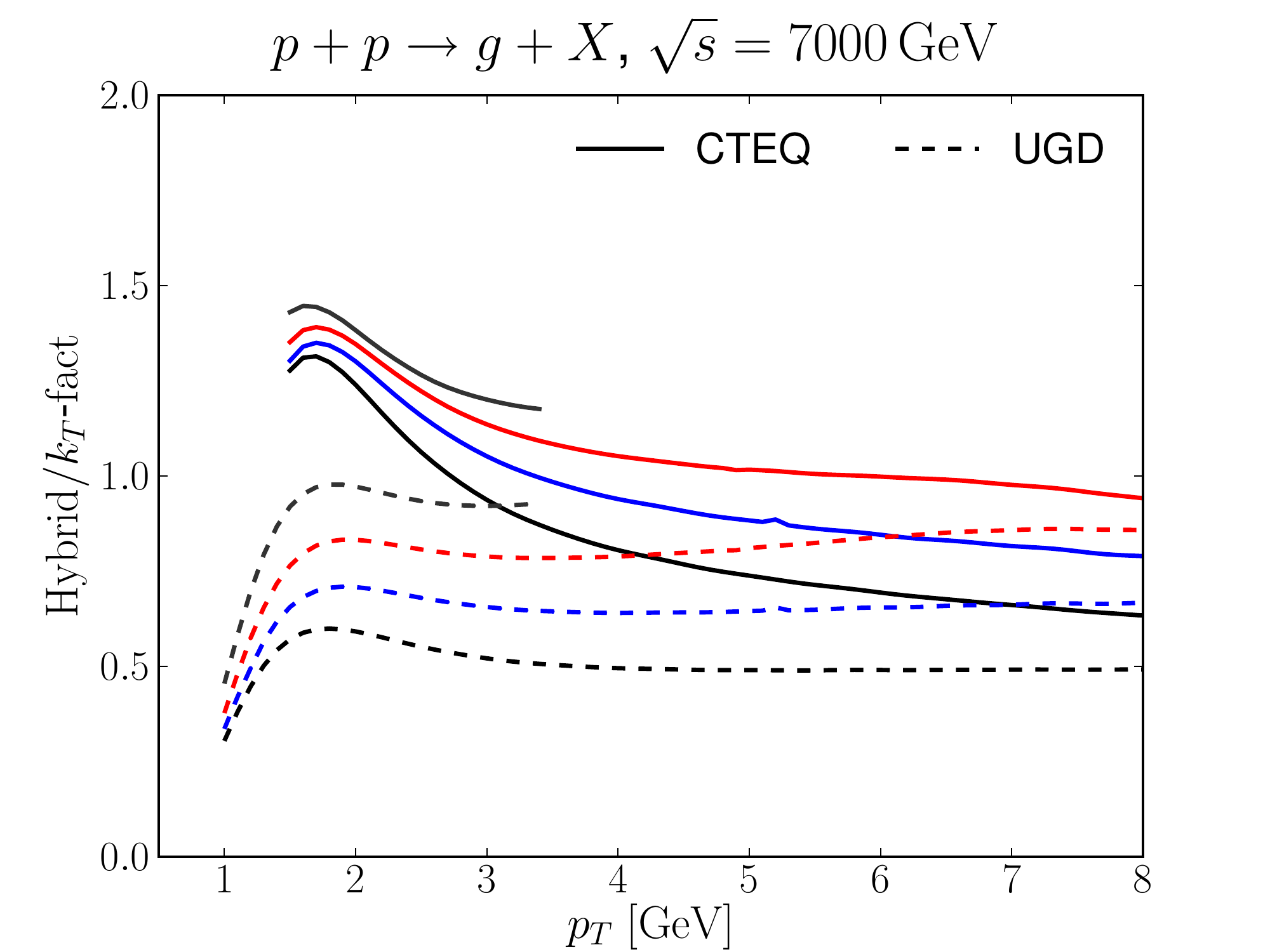}
\end{center}
\caption{
Single inclusive gluon production spectrum at $\sqrt{s}=7000$ GeV obtained by using the hybrid formalism and CTEQ (solid lines) or UGD (dashed lines) parton gluon distribution functions normalized by the corresponding spectrum obtained by using the $k_T$ factorization. The rapidities are, from bottom to top, $y=0,1,2,3$. The results are shown in the kinematical region where $x<10^{-2}$.
}
\label{fig:parton}
\end{figure}

\begin{figure}[tb]
\begin{center}
\includegraphics[width=0.49\textwidth]{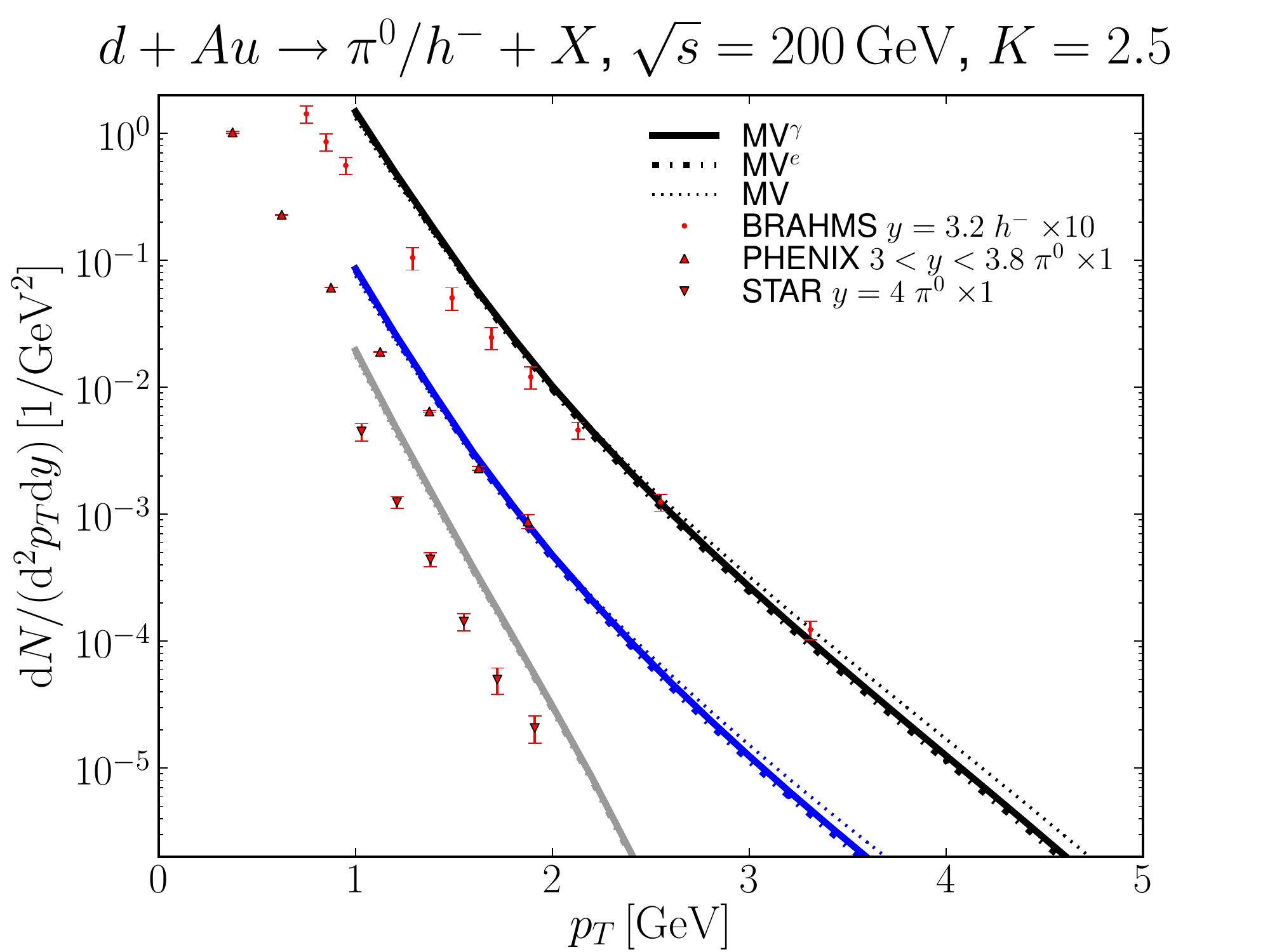}
\end{center}
\caption{
Single inclusive $\pi^0$ and negative hadron production at $\sqrt{s}=200$ GeV d+Au collisions compared with BRAHMS~\cite{Arsene:2004ux}, STAR~\cite{Adams:2006uz} and PHENIX~\cite{Meredith:2011nla} data.\\
}\label{fig:rhic_dau_yield}
\end{figure}

\begin{figure}[tb]
\begin{center}
\includegraphics[width=0.49\textwidth]{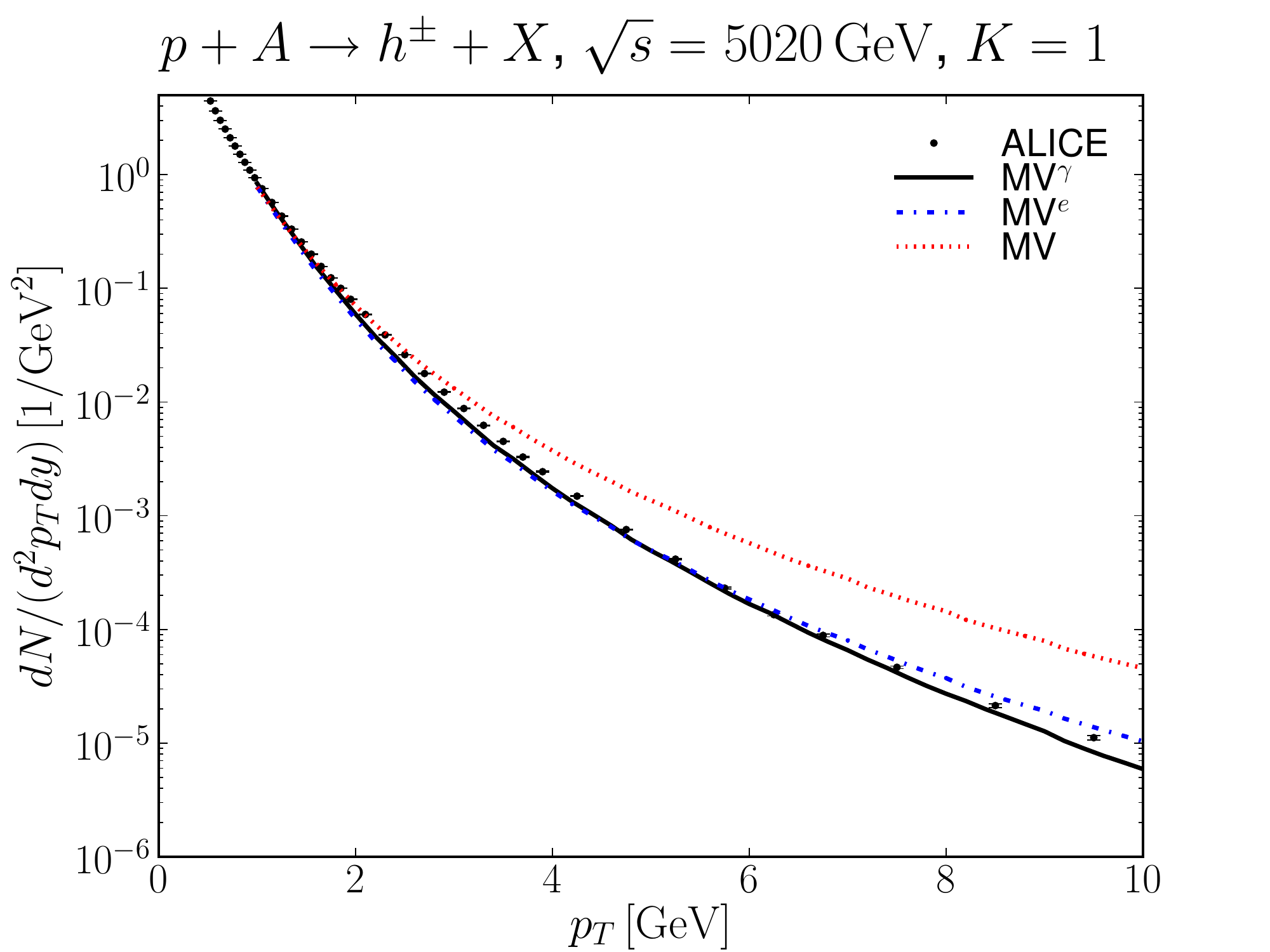}
\end{center}
\caption{
Single inclusive charged hadron production in minimum bias p+Pb collisions at $\sqrt{s}=5020$ GeV computed using the $k_T$ factorization and compared with ALICE data \cite{Abelev:2012mj}.\\
}\label{fig:alice_pa_yield}
\end{figure}

\begin{figure}[tb]
\begin{center}
\includegraphics[width=0.49\textwidth]{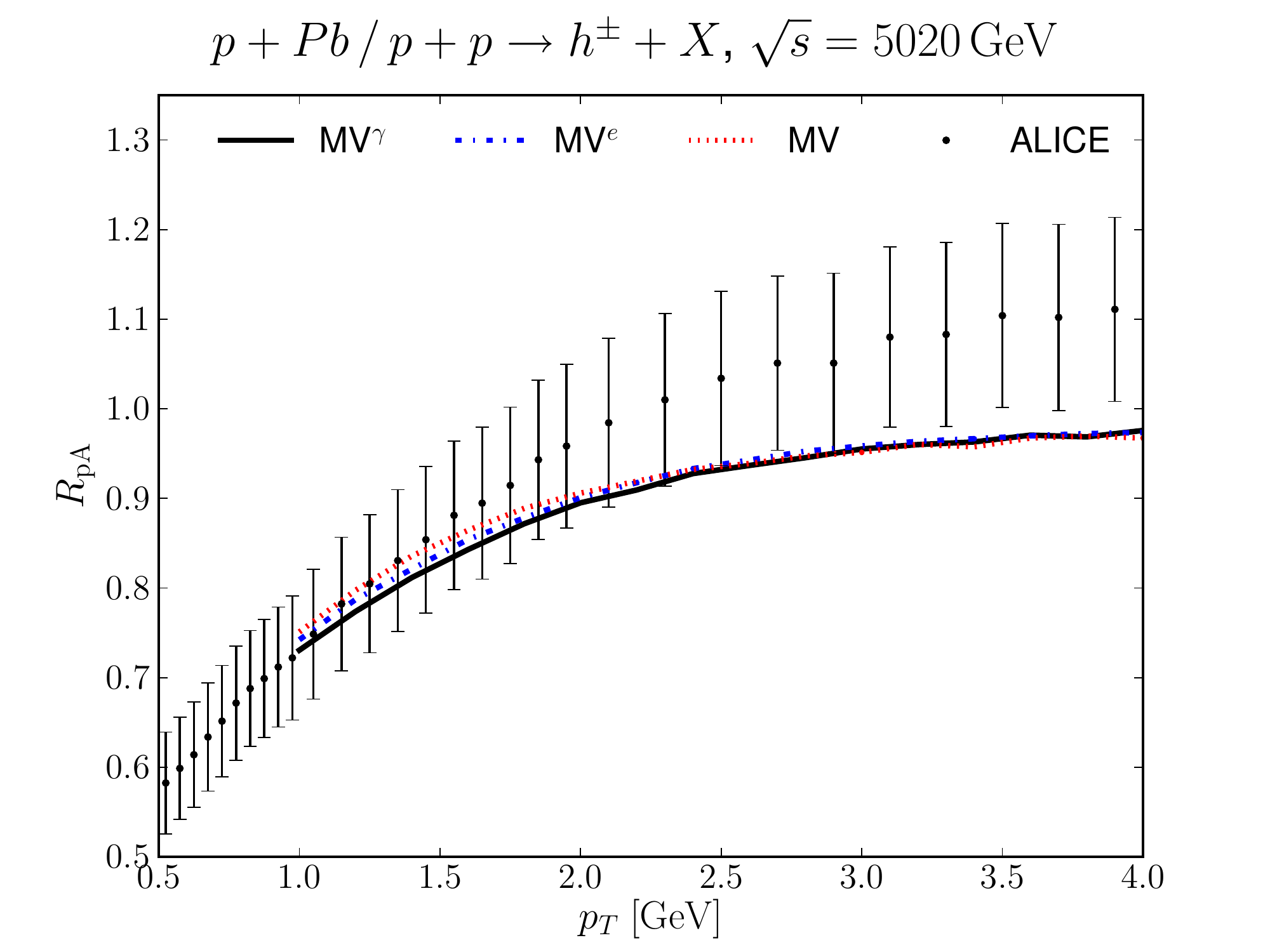}
\end{center}
\caption{
Minimum bias nuclear suppression factor $R_{pA}(y=0)$ at different centrality classes computed using $k_T$ factorization and MV$^\gamma$, MV$^e$ and MV model initial conditions compared with the minimum bias ALICE data \cite{Abelev:2012mj} at smallest $p_T$ region.
}\label{fig:alice_rpa_dipole}
\end{figure}

\begin{figure}[tb]
\begin{center}
\includegraphics[width=0.49\textwidth]{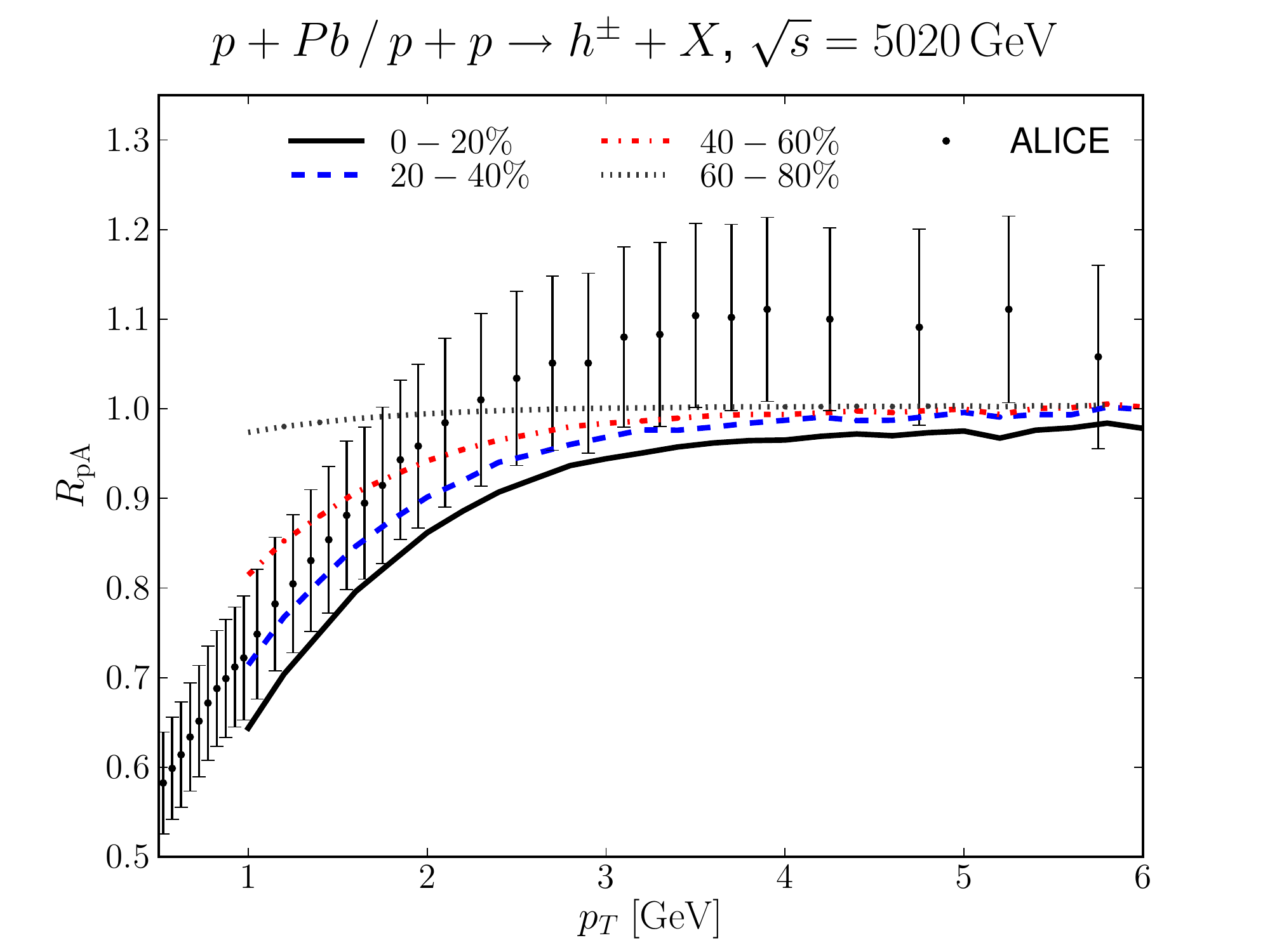}
\end{center}
\caption{
Centrality dependence of $R_{pA}(y=0)$ computed using $k_T$ factorization and MV$^e$ initial condition compared with the ALICE data \cite{Abelev:2012mj}.
}\label{fig:alice_rpa_c}
\end{figure}

\begin{figure}[tb]
\begin{center}
\includegraphics[width=0.49\textwidth]{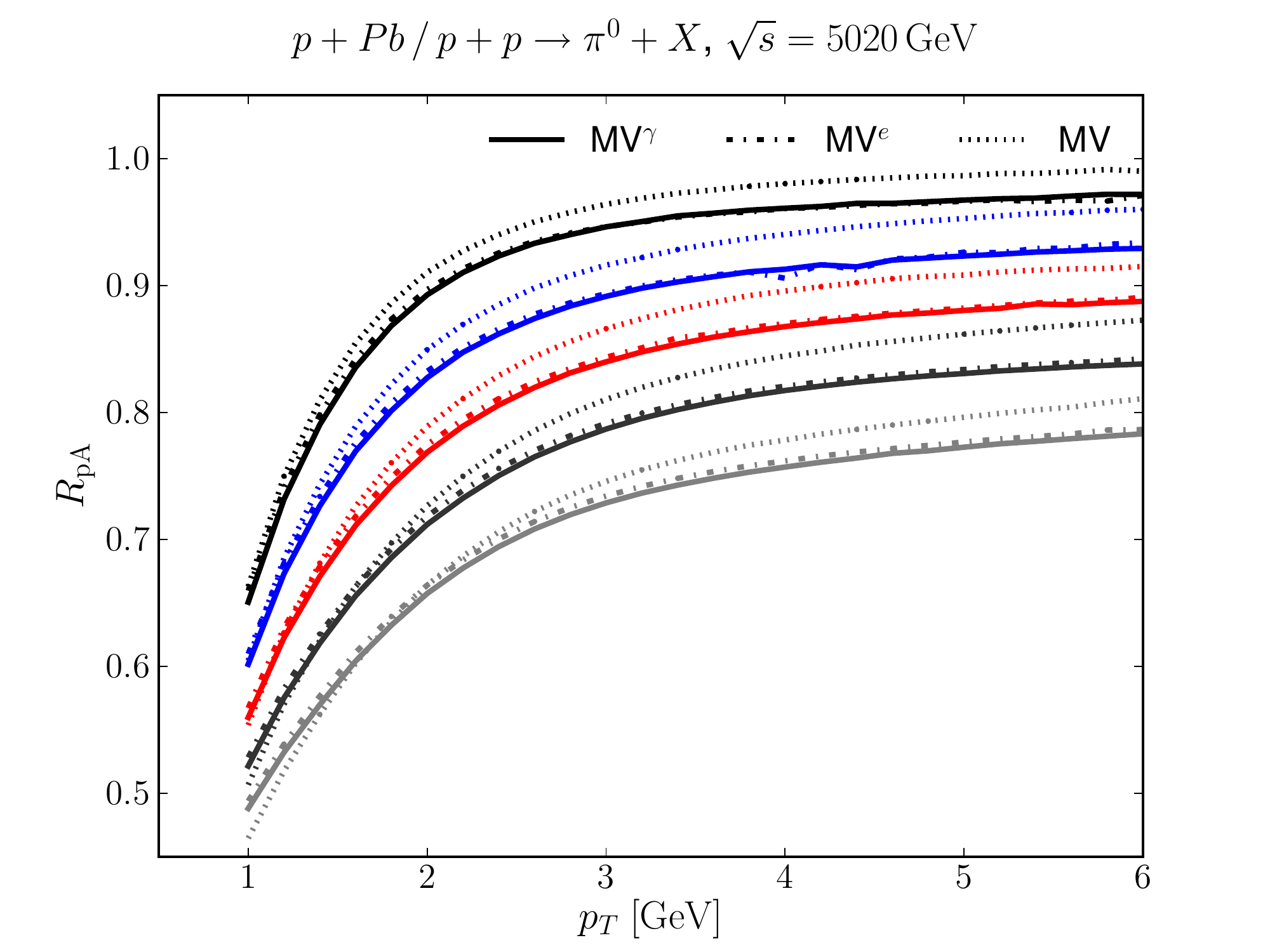}
\end{center}
\caption{
Rapidity dependence of the nuclear modification factor at rapidities $y=2,3,4,5,6$ (from top to bottom) using the MV, MV$^\gamma$ and MV$^e$ initial conditions.\\
}\label{fig:rpa_ydep_minbias}
\end{figure}

\begin{figure}[tb]
\begin{center}
\includegraphics[width=0.49\textwidth]{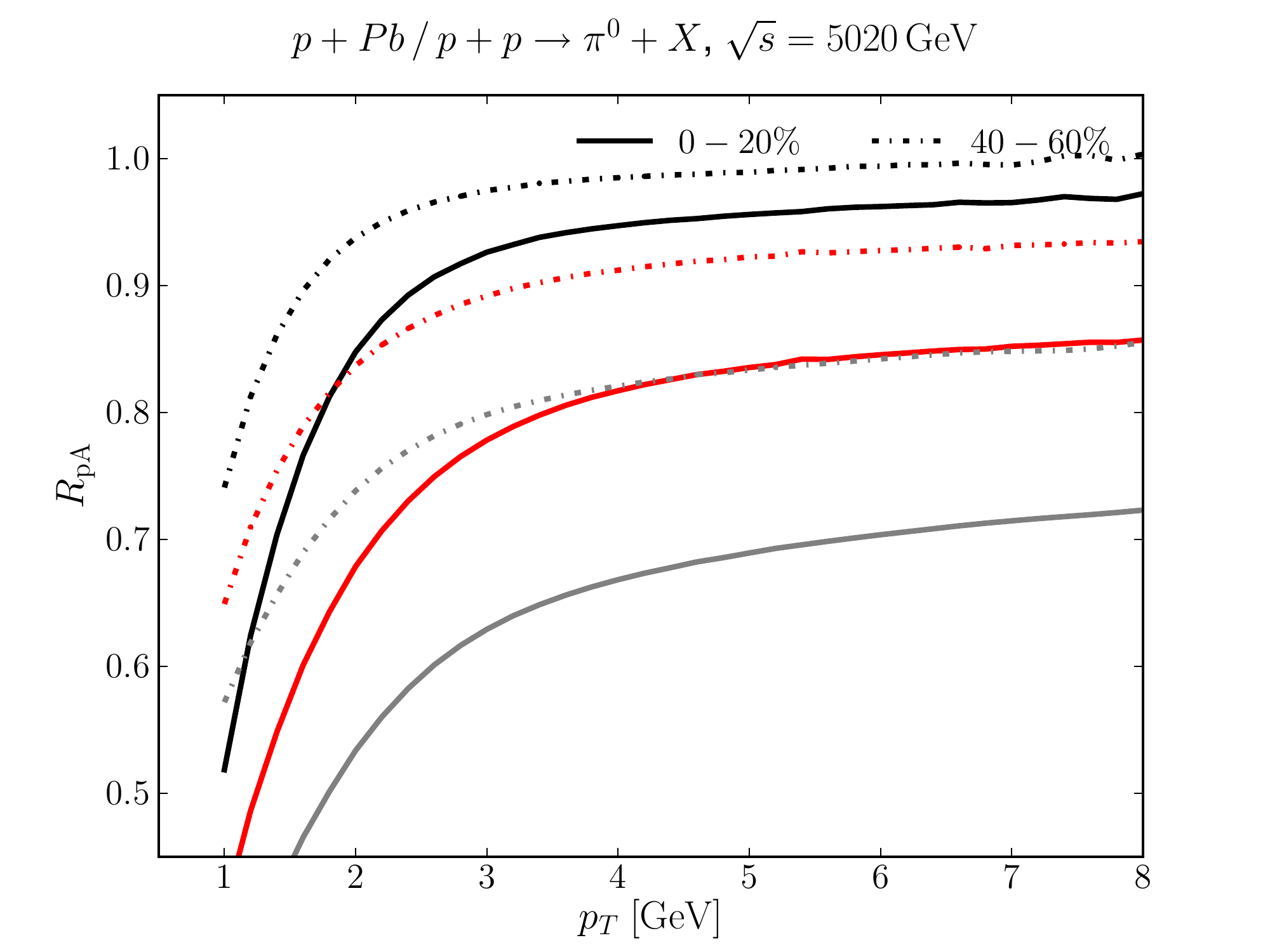}
\end{center}
\caption{
Nuclear modification factor at rapidities $y=2,4,6$ (from top to bottom) using the MV$^\gamma$ initial condition at centrality classes $0-20\%$ and $40-60\%$.}
\label{fig:rpa_ydep_centrality}
\end{figure}

\begin{figure}[tb]
\begin{center}
\includegraphics[width=0.49\textwidth]{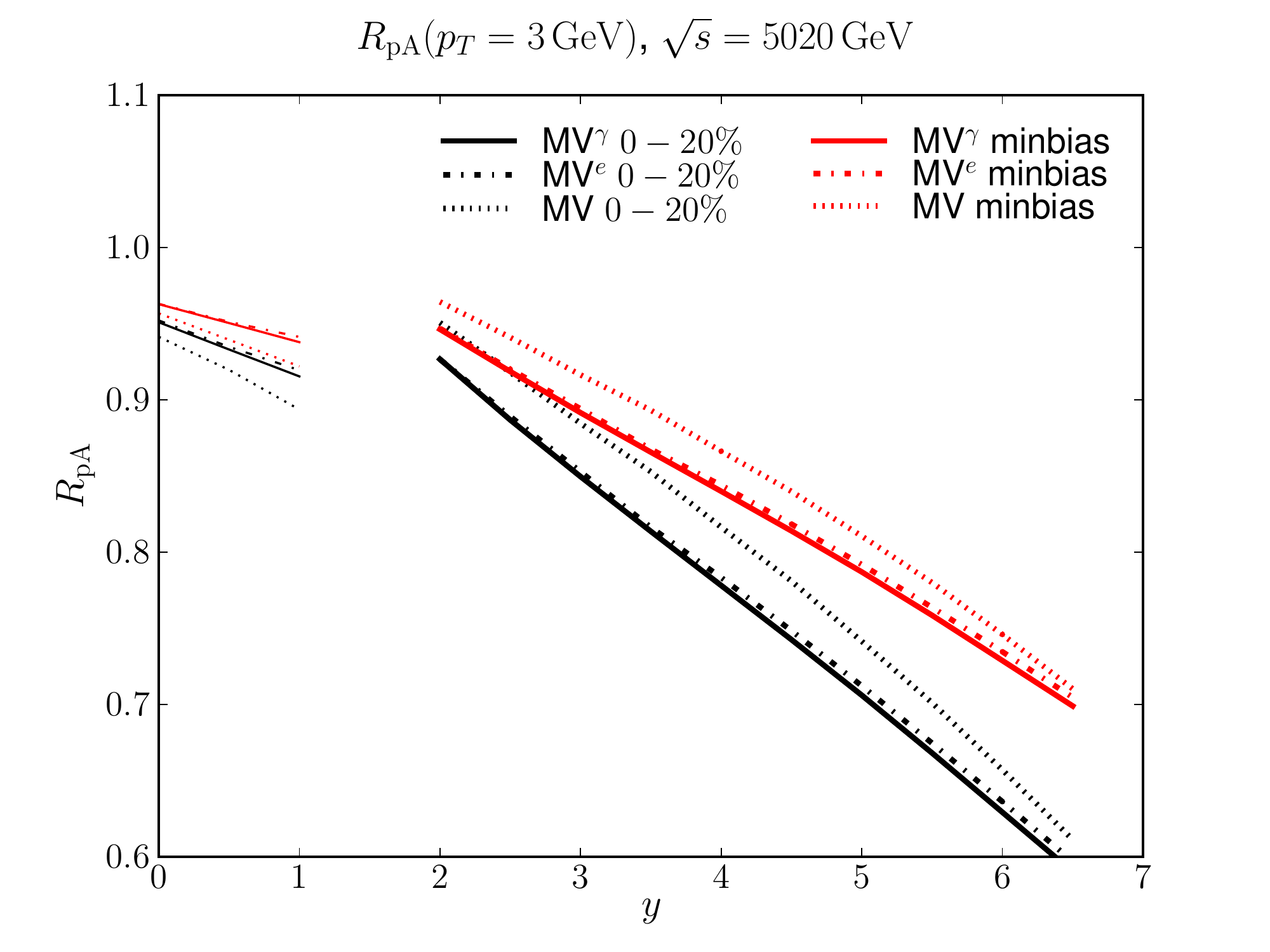}
\end{center}
\caption{
Rapidity and centrality dependence of the nuclear modification factor in neutral pion production in $0-20\%$ most central (solid lines) and minimum bias collisions using MV, MV$^\gamma$ and MV$^e$ initial conditions. Thin lines at $y\le1$ are computed using the $k_T$ factorization and thick lines at $y\ge 2$ using the hybrid formalism.
}\label{fig:rpa_ydep_fixedpt}
\end{figure}

\begin{figure}[tb]
\begin{center}
\includegraphics[width=0.49\textwidth]{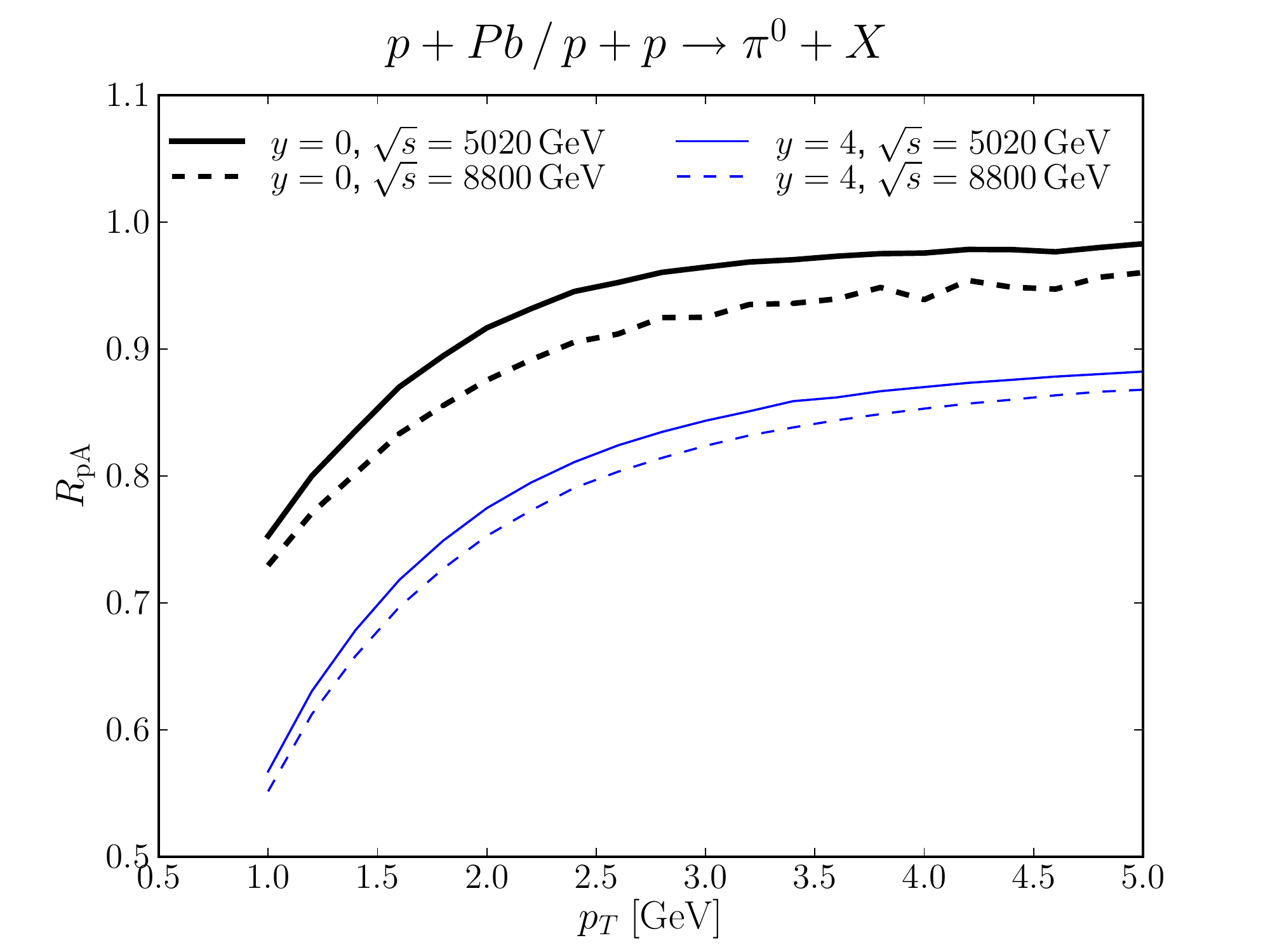}
\end{center}
\caption{
Center of mass energy ($\sqrt{s}$) dependence of the nuclear modification factor in neutral pion production in minimum-bias p+Pb collisions. computed using the MV$^e$ initial condition. The results at midrapidity $y=0$ are computed using the $k_T$ factorisation and at $y=4$ the hybrid formalism is used.
}\label{fig:rpa_sqrts}
\end{figure}

In Fig. \ref{fig:rhic_pp_yield} we show the single inclusive $\pi^0$ and negative hadron yields computed using the hybrid formalism at $\sqrt{s}=200$ GeV and compare with the experimental data from RHIC \cite{Adams:2006uz,Adare:2011sc,Arsene:2004ux}. As an initial condition for the BK evolution we use MV, MV$^\gamma$ and MV$^e$ fits. We recall that all fits, especially MV$^\gamma$ and MV$^e$, give good description of the HERA DIS data (see Fig. \ref{fig:sigmar}). We observe that all initial conditions yield very similar particle spectra, and especially the STAR $\pi^0$ spectra work very well, using 
$K=2.5$. The agreement with BRAHMS and PHENIX data is still reasonably good even though the $p_T$ slope is not exactly correct. 

Fig. \ref{fig:tevatron_yield} shows a comparison with the CDF charged hadron data~\cite{Aaltonen:2009ne} at $\sqrt{s}=1960$ GeV computed using $k_T$ factorization. The Tevatron data seems to require a $K$ factor $K\sim 2$, and our calculation slightly overestimates the yield at small $p_T$.  The standard MV model does not any more give a reasonable description of the data, whereas the MV$^\gamma$ and MV$^e$ models are in good agreement, still giving a slightly wrong slope at small $p_T$.

The single inclusive $\pi^0$ and charged hadron yields at $\sqrt{s}=7000$ GeV computed using $k_T$-factorization and compared with ALICE~\cite{Abelev:2012cn}  and CMS~\cite{Khachatryan:2010us} data are shown in Fig. \ref{fig:alice_cms_pp_yield}. Both MV$^\gamma$ and MV$^e$ models describe the data well without any additional $K$ factor\footnote{As the Tevatron data seems to require a $K$ factors $\sim 2$ and no such a factor is needed for the LHC, it seems that the required $K$ factor decreases as a function of energy. A similar result for LO pQCD calculations was found in Ref. \cite{Eskola:2002kv}.}. The pure MV model gives a too hard spectrum, similarly as with the Tevatron data.

In order to study the sensitivity to different ingredients we compute also the neutral pion spectrum at LHC energies using both $k_T$-factorization and the hybrid formalism using the MV$^e$ initial condition. The results are shown in Fig. \ref{fig:alice_pi0_models}. With the hybrid formalism we use both CTEQ parton distribution functions and a gluon distribution obtained by integrating the unintegrated gluon distribution, see \eq \eqref{eq:xg}. We also compare the DSS and KKP fragmentation functions (we use consistently only LO distributions in this work). The results are scaled by a $K$ factor which is chosen to fit the data around $\pt \approx 2\,\mathrm{GeV}$. We observe that apart from the different overall normalization the different model combinations give very similar spectra, the hybrid formalism with unintegrated gluon distribution PDF deviating slightly from the data at large $|\pt|$. Notice that when using the UGD parton distribution function we can only compute the gluon channel as we have no straightforward way to obtain the quark distribution from the gluon density. Similar conclusions are obtained when the analysis is performed with NLO PDFs and FFs, when one generally needs slightly larger $K$ factors and the $\pt$ slope obtained using the hybrid formalism is slightly worse.

We can conclude from Fig. \ref{fig:alice_pi0_models} that the absolute normalization 
depends strongly on the choice of hybrid vs. $k_T$-factorized formalisms, and also the 
fragmentation function set used. 
The $\pt$ slope, on the other hand, is a more solid prediction of the BK evolution. This is easily understood as the $\pt$ is directly related to the probed Bjorken $x$ of the target, as (neglecting fragmentation) $x \sim p_T e^{-y}/\sqrt{s}$, thus the $\pt$ dependence is given by the BK equation.

In order to study the differences between the $k_T$-factorization and the hybrid formalisms we plot in Fig. \ref{fig:parton} the parton level gluon production yield at rapidities $y=0,1,2$ and $3$ at LHC energies, $\sqrt{s}=7000$ GeV computed using the hybrid formalism (with both CTEQ and UGD parton distribution function) and normalized by the corresponding yield obtained using$k_T$-factorization. We observe that at more forward rapidities, where the saturation scale of one of the protons is relatively small, the hybrid formalism and the $k_T$ factorization are relatively close to each other. This is especially clear when the gluon distribution is computed from the same dipole amplitude (called UGD parton distribution function). We conclude that as the $k_T$-factorization and the hybrid formalisms are relatively close to each other around $y\sim 2$, and as the $k_T$-factorization can not be used in kinematics when the $x$ of one of the protons is large, it is reasonable to switch to the hybrid formalism around $y\sim 2$.


Let us then discuss proton-nucleus and deuteron-nucleus collisions. In Fig. \ref{fig:rhic_dau_yield} we present the single inclusive $\pi^0$ and negative charged hadron production yields in minimum bias deuteron-nucleus collisions at forward rapidities computed using the hybrid formalism with different initial conditions for the BK evolution and compared with the RHIC data~\cite{Arsene:2004ux,Adams:2006uz,Meredith:2011nla}. We use here the same $K$ factor $K=2.5$ that was required to obtain correct normalization with the RHIC pp data. The $p_T$ slopes agree roughly with the data, the agreement being very good with the STAR data. The absolute normalization now works relatively well with the BRAHMS and PHENIX data  (which were underestimated in proton-proton case) whereas the STAR yield is overestimated by a factor of $\sim 2$.

In Fig. \ref{fig:alice_pa_yield} we show the single inclusive charged hadron yield in proton-lead collisions compared with the ALICE data~\cite{Abelev:2012mj}. The conclusion is very similar as for proton-proton collisions (see Fig. \ref{fig:alice_cms_pp_yield}): the pure MV model gives a completely wrong $p_T$ slope, but both MV$^\gamma$ and MV$^e$ models describe the data well.

In Fig. \ref{fig:alice_rpa_dipole} we compare our result for midrapidity minimum bias $R_{pA}$ with ALICE charged hadron measurements~\cite{Abelev:2012mj}. Even though the $p_T$ spectra obtained using the MV$^\gamma$ and MV$^e$ initial conditions are very different from the pure MV model, all three initial conditions yield a very similar $R_{pA}$. Recall that we get exactly $R_{pA}\to 1$ at large $p_T$, consistently with the ALICE result. The centrality dependence of midrapidity $R_{pA}$ is shown in Fig.~\ref{fig:alice_rpa_c}. Here we only compute results using the MV$^e$ initial condition as the results obtained using different initial conditions are basically the same. The centrality dependence is relatively weak, the results start to differ only at most peripheral classes where centrality $\gtrsim 60\%$. Notice that in our calculation we set explicitly $R_{pA}=1$ at centralities $\gtrsim 70\%$, see discussion in Sec. \ref{sec:nuke}.

Let us then present predictions for the future $R_{pA}$ measurements. In Fig. \ref{fig:rpa_ydep_minbias} we show $R_{pA}$ at $\sqrt{s}=5020$ GeV in minimum bias  collisions at forward rapidities. As can be seen from Fig. \ref{fig:parton}, at more forward rapidities the hybrid formalism and the $k_T$-factorization are closer to each other, and we compute the nuclear suppression factors using the hybrid formalism at $y\ge 2$. Note that it is not possible to use $k_T$ factorization in this kinematical region as one also has a relatively large component from the large-$x$ part of the proton. We compute $R_{pA}$ using the MV, MV$^\gamma$ and MV$^e$ initial conditions, and we see that their difference remains small at all rapidities. The centrality dependence of $R_{pA}$ is shown in Fig. \ref{fig:rpa_ydep_centrality}. At $y=2$ most central and peripheral collisions give a similar $R_{pA}$, and the difference between the two centrality classes increases when we move to more forward rapidities.  

We observe slightly less suppression than obtained in Ref. \cite{Albacete:2012xq} where the saturation scale of the nucleus is computed in a Monte Carlo Glauber model. We conjecture that this is due to the fact that since $\sigma_0/2 < 42$ mb, the nuclear $Q_s$ is smaller than assumed in Ref. \cite{Albacete:2012xq}. The evolution speed obtained using the Monte Carlo method is very similar than what is obtained in this work.

In order to further demonstrate the evolution speed of the nuclear modification factor we plot $R_{pA}(\pt=3\,\mathrm{GeV})$ for neutral pion production at LHC energies in Fig. \ref{fig:rpa_ydep_fixedpt} in most central and minimum bias collisions. We compute $R_{pA}$ close to midrapidity using $k_T$-factorization and at forward rapidites using the hybrid formalism, where we use CTEQ the parton distribution function and also include the quark initated channel. Thus the obtained curve is not exactly continuous. The evolution speed close to midrapidity (where $k_T$-factorization should be valid) is slightly slower than at more forward rapidities where the hybrid formalism is more reliable.
The MV model initial condition gives a slightly different result than the MV$^\gamma$ and MV$^e$ models, and all dipole models give basically the same evolution speed. Thus $R_{pA}$ is not sensitive to the details of the initial dipole amplitude, and the evolution speed of $R_{pA}$ is driven by the BK evolution. The centrality and especially rapidity evolution speed is significantly faster than in a NLO pQCD calculation using the EPS09s nuclear parton distribution functions~\cite{Helenius:2012wd,helenius2013rapidity}.

Finally we demonstrate the energy dependence of the nuclear modification factor by showing in Fig. \ref{fig:rpa_sqrts} $R_{pA}$ at midrapidity and at $y=4$ in minimum-bias p+Pb collisions. Increasing the energy from the current $\sqrt{s}=5020$ GeV to the design energy $\sqrt{s}=8800$ GeV (where we use $\sigma_\text{inel}=75$ mb~\cite{Antchev:2011vs}) does not change $R_{pA}$ significantly, and we get midrapidity $R_{pA}\to 1$ at large $p_T$ at all $\sqrt{s}$ as discussed in Sec. \ref{sec:nuke}. The result differs significantly from the corresponding prediction shown in Ref. \cite{Tribedy:2011aa} where a much faster energy evolution was predicted.

\section{Conclusions}
\label{sec:conclusions}
Taking only input from electron-proton deep inelastic scattering and standard nuclear geometry we compute single inclusive hadron production in proton-proton and proton-nucleus collisions from the Color Glass Condensate framework. We observe that in order to obtain a consistent description of all the single inclusive data one has to modify the MV model, which is used as an initial condition for the BK evolution. We show that while a modification is required, one does not have to introduce an anomalous dimension $\gamma>1$, but instead it is enough to take the infrared cutoff in the MV model to be a fit parameter. Using the pure MV model (without anomalous dimension or modification to the infrared cutoff) one also obtains a reasonable good description of the HERA and RHIC data, but Tevatron and LHC proton-proton data clearly favour models with an anomalous dimension or scaling of the infrared cutoff parameter.

We obtain a good description of the available proton-nucleus and deuteron-nucleus data, the absolute normalizations of the RHIC results being difficult to reproduce simultaneously, as is the case also with the RHIC forward proton-proton data. We obtain exactly $R_{pA}\to 1$ at large $\pt$ which is a natural requirement and consistent with the available ALICE data, and follows directly from our consistent treatment of the difference between the proton transverse areas measured in DIS and the inelastic proton-proton cross section. We present predictions for the future forward $R_{pA}$ measurements. Especially we find that the rapidity evolution of the $R_{pA}$ at fixed $\pt$ is a solid prediction of the CGC, given by the BK equation.

\appendix
\section{Optical glauber}
\label{glauber}
Let us briefly specify the optical Glauber model used here. In a proton-nucleus
collision at an impact parameter $\bt$ the number of binary collisions is given
by\begin{equation}
	N_\text{bin} = A T_A(\bt) \sigma_\text{inel}
\end{equation}
where $\sigma_\text{inel}$ is the total inelastic nucleon-nucleon
cross section and  $T_A(\bt)$ is the transverse thickness function of the 
nucleus obtained by integrating the Woods-Saxon distribution 
\begin{equation}
	\rho_A(\bt, z) = \frac{n}{1+\exp \left[ \frac{\sqrt{\bt^2 + z^2}+R_A}{d} \right]}
\end{equation}
over the longitudinal distance $z$. The parameters are $d=0.54\,\mathrm{fm}$ and
$R_A=(1.12A^{1/3}-0.86A^{-1/3})\,\mathrm{fm}$. The distribution is normalized
to unity, $\int \der^2 \bt \der z \rho_A(\bt, z) = 1$.

The probability for having an inelastic collision is
\begin{equation}
	p(\bt) \approx 1- e^{-A T_A(\bt) \sigma_\text{inel}},
\end{equation}
and the total inelastic proton-nucleus cross section is then
\begin{equation}
	\sigma_\text{inel}^{pA} = \int \der^2 \bt \, p(\bt).
\end{equation}

A centrality class $(c_1-c_2)\%$ corresponds to impact parameter interval $[b_1,b_2]$ for which
\begin{equation}
	(c_1-c_2)\% = \frac{1}{\sigma_\text{inel}^{pA}} \int_{b_1}^{b_2} \der^2 \bt p(\bt).
\end{equation}
The $(0-c)\%$ most central collisions give $c\ \%$ of the total inelastic 
proton-nucleus cross section.

The average number of binary collisions in certain impact parameter class is
\begin{equation}
	\langle N_\text{bin} \rangle_{b_1,b_2} = \frac{\int_{b_1}^{b_2} \der^2 \bt N_\text{bin}(\bt)}{\int_{b_1}^{b_2} \der^2 \bt p(\bt)},
\end{equation}
where the denominator is $(c_1-c_2)\%$ of the total inelastic proton-nucleus 
cross section $\sigma_\text{inel}^{pA}$.
The centrality classes and the corresponding values of $N_\text{bin}$ for RHIC
and LHC energies are shown in Tables \ref{tab:centrality_rhic} and
\ref{tab:centrality_lhc}.

The particle yield in a centrality class is computed as
\begin{equation}
	\frac{\der N}{\der y \der^2 \kt} = \frac{ \int \der^2 \bt \frac{\der N(\bt)}{\der y \der^2 \kt} } {\int \der^2 \bt \, p(\bt)},
\end{equation}
where integration limits are set according to the corresponding centrality 
class.

\begin{table}
\begin{tabular}{|l||r|r|r|}
\hline
Bin & $b_1$ [fm] & $b_2$ [fm] & $\langle N_\text{bin}/2 \rangle$ \\
\hline\hline
0--20\% & 0.0 &  3.26 & 8.45 \\ 
20--40\% & 3.26 & 4.62 & 6.95 \\
40--60\% & 4.62 & 5.66 & 5.03 \\
60--80\% & 5.66 & 6.61 & 2.89 \\ 
0-100\% & 0 & & 4.95 \\
\hline
\end{tabular}
\caption{Impact parameter intervals and number of binary collisions in deuteron-gold collisions for different centrality classes at RHIC energies, where
$\sqrt{s}=200$ GeV and $\sigma_\text{inel}=42$ mb. We assume that assume that deuteron consist of two independent nucleons, giving $N_\text{bin}^{dA}=2N_\text{bin}^{pA}$.
}
\label{tab:centrality_rhic}
\end{table}

\begin{table}
\begin{tabular}{|l||r|r|r|}
\hline
Bin & $b_1$ [fm] & $b_2$ [fm] & $\langle N_\text{bin} \rangle$ \\
\hline\hline
0--20\% & 0.0 &  3.47 & 14.24 \\ 
20--40\% & 3.47 & 4.91 & 11.41 \\
40--60\% & 4.91 & 6.01 & 7.66 \\
60--80\% & 6.01 & 6.99 & 3.68 \\
0--100\% & 0.0 & & 7.69 \\
\hline
\end{tabular}
\caption{Impact parameter intervals and number of binary collisions in proton-lead collisions at LHC energies, where $\sqrt{s}=5020$ GeV and $\sigma_\text{inel}=70$ mb. 
}
\label{tab:centrality_lhc}
\end{table}

\section*{Acknowledgements}
We thank I. Helenius and L. Korkeala for discussions and M. Chiu for providing
us the PHENIX $\pi^0$ yield in proton-proton collisions.
This work has been supported by the Academy of Finland, projects 133005, 
267321, 273464 and by computing resources from CSC -- IT Center for 
Science in Espoo, Finland. H.M. is supported by the Graduate School of 
Particle and Nuclear Physics.

\bibliography{refs}

\ifx\mcitethebibliography\mciteundefinedmacro
\PackageError{JHEP-2modM.bst}{mciteplus.sty has not been loaded}
{This bibstyle requires the use of the mciteplus package.}\fi
\providecommand{\href}[2]{#2}
\begingroup\raggedright\begin{mcitethebibliography}{10}

\bibitem{Albacete:2010sy}
J.~L. Albacete, N.~Armesto, J.~G. Milhano, P.~Quiroga-Arias and C.~A. Salgado,
  \mbox{}  \href{http://dx.doi.org/10.1140/epjc/s10052-011-1705-3}{{\em Eur.
  Phys. J.} {\bf C71} (2011) 1705} [\href{http://arXiv.org/abs/1012.4408}{{\tt
  arXiv:1012.4408 [hep-ph]}}]\relax
\mciteBstWouldAddEndPuncttrue
\mciteSetBstMidEndSepPunct{\mcitedefaultmidpunct}
{\mcitedefaultendpunct}{\mcitedefaultseppunct}\relax
\EndOfBibitem
\bibitem{Rezaeian:2012ji}
A.~H. Rezaeian, M.~Siddikov, M.~Van~de Klundert and R.~Venugopalan,  \mbox{}
  {\em Phys.Rev.} {\bf D87} (2013) 034002
  [\href{http://arXiv.org/abs/1212.2974}{{\tt arXiv:1212.2974 [hep-ph]}}]\relax
\mciteBstWouldAddEndPuncttrue
\mciteSetBstMidEndSepPunct{\mcitedefaultmidpunct}
{\mcitedefaultendpunct}{\mcitedefaultseppunct}\relax
\EndOfBibitem
\bibitem{Rezaeian:2013tka}
A.~H. Rezaeian and I.~Schmidt,  \mbox{}
  \href{http://arXiv.org/abs/1307.0825}{{\tt arXiv:1307.0825 [hep-ph]}}\relax
\mciteBstWouldAddEndPuncttrue
\mciteSetBstMidEndSepPunct{\mcitedefaultmidpunct}
{\mcitedefaultendpunct}{\mcitedefaultseppunct}\relax
\EndOfBibitem
\bibitem{Tribedy:2011aa}
P.~Tribedy and R.~Venugopalan,  \mbox{}
  \href{http://dx.doi.org/10.1016/j.physletb.2012.02.047,
  10.1016/j.physletb.2012.12.004}{{\em Phys. Lett.} {\bf B710} (2012) 125}
  [\href{http://arXiv.org/abs/1112.2445}{{\tt arXiv:1112.2445 [hep-ph]}}]\relax
\mciteBstWouldAddEndPuncttrue
\mciteSetBstMidEndSepPunct{\mcitedefaultmidpunct}
{\mcitedefaultendpunct}{\mcitedefaultseppunct}\relax
\EndOfBibitem
\bibitem{Albacete:2012xq}
J.~L. Albacete, A.~Dumitru, H.~Fujii and Y.~Nara,  \mbox{}
  \href{http://dx.doi.org/10.1016/j.nuclphysa.2012.09.012}{{\em Nucl. Phys.}
  {\bf A897} (2013) 1} [\href{http://arXiv.org/abs/1209.2001}{{\tt
  arXiv:1209.2001 [hep-ph]}}]\relax
\mciteBstWouldAddEndPuncttrue
\mciteSetBstMidEndSepPunct{\mcitedefaultmidpunct}
{\mcitedefaultendpunct}{\mcitedefaultseppunct}\relax
\EndOfBibitem
\bibitem{Rezaeian:2012ye}
A.~H. Rezaeian,  \mbox{}
  \href{http://dx.doi.org/10.1016/j.physletb.2012.11.066}{{\em Phys.Lett.} {\bf
  B718} (2013) 1058} [\href{http://arXiv.org/abs/1210.2385}{{\tt
  arXiv:1210.2385 [hep-ph]}}]\relax
\mciteBstWouldAddEndPuncttrue
\mciteSetBstMidEndSepPunct{\mcitedefaultmidpunct}
{\mcitedefaultendpunct}{\mcitedefaultseppunct}\relax
\EndOfBibitem
\bibitem{Lappi:2012nh}
T.~Lappi and H.~M{\"a}ntysaari,  \mbox{}
  \href{http://dx.doi.org/10.1016/j.nuclphysa.2013.03.017}{{\em Nucl. Phys.}
  {\bf A908} (2013) } [\href{http://arXiv.org/abs/1209.2853}{{\tt
  arXiv:1209.2853 [hep-ph]}}]\relax
\mciteBstWouldAddEndPuncttrue
\mciteSetBstMidEndSepPunct{\mcitedefaultmidpunct}
{\mcitedefaultendpunct}{\mcitedefaultseppunct}\relax
\EndOfBibitem
\bibitem{Albacete:2010pg}
J.~L. Albacete and C.~Marquet,  \mbox{}
  \href{http://dx.doi.org/10.1103/PhysRevLett.105.162301}{{\em Phys. Rev.
  Lett.} {\bf 105} (2010) 162301} [\href{http://arXiv.org/abs/1005.4065}{{\tt
  arXiv:1005.4065 [hep-ph]}}]\relax
\mciteBstWouldAddEndPuncttrue
\mciteSetBstMidEndSepPunct{\mcitedefaultmidpunct}
{\mcitedefaultendpunct}{\mcitedefaultseppunct}\relax
\EndOfBibitem
\bibitem{Stasto:2012ru}
A.~Stasto, B.-W. Xiao and D.~Zaslavsky,  \mbox{}
  \href{http://dx.doi.org/10.1103/PhysRevD.86.014009}{{\em Phys. Rev.} {\bf
  D86} (2012) 014009} [\href{http://arXiv.org/abs/1204.4861}{{\tt
  arXiv:1204.4861 [hep-ph]}}]\relax
\mciteBstWouldAddEndPuncttrue
\mciteSetBstMidEndSepPunct{\mcitedefaultmidpunct}
{\mcitedefaultendpunct}{\mcitedefaultseppunct}\relax
\EndOfBibitem
\bibitem{JalilianMarian:2012bd}
J.~Jalilian-Marian and A.~H. Rezaeian,  \mbox{}
  \href{http://dx.doi.org/10.1103/PhysRevD.86.034016}{{\em Phys. Rev.} {\bf
  D86} (2012) 034016} [\href{http://arXiv.org/abs/1204.1319}{{\tt
  arXiv:1204.1319 [hep-ph]}}]\relax
\mciteBstWouldAddEndPuncttrue
\mciteSetBstMidEndSepPunct{\mcitedefaultmidpunct}
{\mcitedefaultendpunct}{\mcitedefaultseppunct}\relax
\EndOfBibitem
\bibitem{Kowalski:2006hc}
H.~Kowalski, L.~Motyka and G.~Watt,  \mbox{}
  \href{http://dx.doi.org/10.1103/PhysRevD.74.074016}{{\em Phys. Rev.} {\bf
  D74} (2006) 074016} [\href{http://arXiv.org/abs/hep-ph/0606272}{{\tt
  arXiv:hep-ph/0606272}}]\relax
\mciteBstWouldAddEndPuncttrue
\mciteSetBstMidEndSepPunct{\mcitedefaultmidpunct}
{\mcitedefaultendpunct}{\mcitedefaultseppunct}\relax
\EndOfBibitem
\bibitem{Lappi:2013am}
T.~Lappi and H.~M{\"a}ntysaari,  \mbox{}
  \href{http://dx.doi.org/10.1103/PhysRevC.87.032201}{{\em Phys. Rev.} {\bf
  C87} (2013) 032201} [\href{http://arXiv.org/abs/1301.4095}{{\tt
  arXiv:1301.4095 [hep-ph]}}]\relax
\mciteBstWouldAddEndPuncttrue
\mciteSetBstMidEndSepPunct{\mcitedefaultmidpunct}
{\mcitedefaultendpunct}{\mcitedefaultseppunct}\relax
\EndOfBibitem
\bibitem{Lappi:2011ju}
T.~Lappi,  \mbox{}
  \href{http://dx.doi.org/10.1016/j.physletb.2011.08.011}{{\em Phys. Lett.}
  {\bf B703} (2011) 325} [\href{http://arXiv.org/abs/1105.5511}{{\tt
  arXiv:1105.5511 [hep-ph]}}]\relax
\mciteBstWouldAddEndPuncttrue
\mciteSetBstMidEndSepPunct{\mcitedefaultmidpunct}
{\mcitedefaultendpunct}{\mcitedefaultseppunct}\relax
\EndOfBibitem
\bibitem{Schenke:2012wb}
B.~Schenke, P.~Tribedy and R.~Venugopalan,  \mbox{}
  \href{http://dx.doi.org/10.1103/PhysRevLett.108.252301}{{\em Phys. Rev.
  Lett.} {\bf 108} (2012) 252301} [\href{http://arXiv.org/abs/1202.6646}{{\tt
  arXiv:1202.6646 [nucl-th]}}]\relax
\mciteBstWouldAddEndPuncttrue
\mciteSetBstMidEndSepPunct{\mcitedefaultmidpunct}
{\mcitedefaultendpunct}{\mcitedefaultseppunct}\relax
\EndOfBibitem
\bibitem{Gale:2012rq}
C.~Gale, S.~Jeon, B.~Schenke, P.~Tribedy and R.~Venugopalan,  \mbox{}
  \href{http://dx.doi.org/10.1103/PhysRevLett.110.012302}{{\em Phys. Rev.
  Lett.} {\bf 110} (2013) 012302} [\href{http://arXiv.org/abs/1209.6330}{{\tt
  arXiv:1209.6330 [nucl-th]}}]\relax
\mciteBstWouldAddEndPuncttrue
\mciteSetBstMidEndSepPunct{\mcitedefaultmidpunct}
{\mcitedefaultendpunct}{\mcitedefaultseppunct}\relax
\EndOfBibitem
\bibitem{Balitsky:2008zza}
I.~Balitsky and G.~A. Chirilli,  \mbox{}
  \href{http://dx.doi.org/10.1103/PhysRevD.77.014019}{{\em Phys. Rev.} {\bf
  D77} (2008) 014019} [\href{http://arXiv.org/abs/0710.4330}{{\tt
  arXiv:0710.4330 [hep-ph]}}]\relax
\mciteBstWouldAddEndPuncttrue
\mciteSetBstMidEndSepPunct{\mcitedefaultmidpunct}
{\mcitedefaultendpunct}{\mcitedefaultseppunct}\relax
\EndOfBibitem
\bibitem{Chirilli:2012jd}
G.~A. Chirilli, B.-W. Xiao and F.~Yuan,  \mbox{}
  \href{http://dx.doi.org/10.1103/PhysRevD.86.054005}{{\em Phys. Rev.} {\bf
  D86} (2012) 054005} [\href{http://arXiv.org/abs/1203.6139}{{\tt
  arXiv:1203.6139 [hep-ph]}}]\relax
\mciteBstWouldAddEndPuncttrue
\mciteSetBstMidEndSepPunct{\mcitedefaultmidpunct}
{\mcitedefaultendpunct}{\mcitedefaultseppunct}\relax
\EndOfBibitem
\bibitem{Stasto:2013cha}
A.~M. Stasto, B.-W. Xiao and D.~Zaslavsky,  \mbox{}
  \href{http://arXiv.org/abs/1307.4057}{{\tt arXiv:1307.4057 [hep-ph]}}\relax
\mciteBstWouldAddEndPuncttrue
\mciteSetBstMidEndSepPunct{\mcitedefaultmidpunct}
{\mcitedefaultendpunct}{\mcitedefaultseppunct}\relax
\EndOfBibitem
\bibitem{Balitsky:1995ub}
I.~Balitsky,  \mbox{}
  \href{http://dx.doi.org/10.1016/0550-3213(95)00638-9}{{\em Nucl. Phys.} {\bf
  B463} (1996) 99} [\href{http://arXiv.org/abs/hep-ph/9509348}{{\tt
  arXiv:hep-ph/9509348}}]\relax
\mciteBstWouldAddEndPuncttrue
\mciteSetBstMidEndSepPunct{\mcitedefaultmidpunct}
{\mcitedefaultendpunct}{\mcitedefaultseppunct}\relax
\EndOfBibitem
\bibitem{Kovchegov:1999yj}
Y.~V. Kovchegov,  \mbox{}
  \href{http://dx.doi.org/10.1103/PhysRevD.60.034008}{{\em Phys. Rev.} {\bf
  D60} (1999) 034008} [\href{http://arXiv.org/abs/hep-ph/9901281}{{\tt
  arXiv:hep-ph/9901281 [hep-ph]}}]\relax
\mciteBstWouldAddEndPuncttrue
\mciteSetBstMidEndSepPunct{\mcitedefaultmidpunct}
{\mcitedefaultendpunct}{\mcitedefaultseppunct}\relax
\EndOfBibitem
\bibitem{Kovchegov:1999ua}
Y.~V. Kovchegov,  \mbox{}
  \href{http://dx.doi.org/10.1103/PhysRevD.61.074018}{{\em Phys. Rev.} {\bf
  D61} (2000) 074018} [\href{http://arXiv.org/abs/hep-ph/9905214}{{\tt
  arXiv:hep-ph/9905214}}]\relax
\mciteBstWouldAddEndPuncttrue
\mciteSetBstMidEndSepPunct{\mcitedefaultmidpunct}
{\mcitedefaultendpunct}{\mcitedefaultseppunct}\relax
\EndOfBibitem
\bibitem{Aaron:2009aa}
{\bf H1 and ZEUS} collaboration, F.~Aaron {\em et.~al.},  \mbox{}
  \href{http://dx.doi.org/10.1007/JHEP01(2010)109}{{\em JHEP} {\bf 1001} (2010)
  109} [\href{http://arXiv.org/abs/0911.0884}{{\tt arXiv:0911.0884
  [hep-ex]}}]\relax
\mciteBstWouldAddEndPuncttrue
\mciteSetBstMidEndSepPunct{\mcitedefaultmidpunct}
{\mcitedefaultendpunct}{\mcitedefaultseppunct}\relax
\EndOfBibitem
\bibitem{Kovchegov:2012mbw}
Y.~V. Kovchegov and E.~Levin, {\em {Quantum chromodynamics at high energy}}.
\newblock Cambridge University Press, 2012\relax
\mciteBstWouldAddEndPuncttrue
\mciteSetBstMidEndSepPunct{\mcitedefaultmidpunct}
{\mcitedefaultendpunct}{\mcitedefaultseppunct}\relax
\EndOfBibitem
\bibitem{Balitsky:2006wa}
I.~Balitsky,  \mbox{}  \href{http://dx.doi.org/10.1103/PhysRevD.75.014001}{{\em
  Phys. Rev.} {\bf D75} (2007) 014001}
  [\href{http://arXiv.org/abs/hep-ph/0609105}{{\tt arXiv:hep-ph/0609105
  [hep-ph]}}]\relax
\mciteBstWouldAddEndPuncttrue
\mciteSetBstMidEndSepPunct{\mcitedefaultmidpunct}
{\mcitedefaultendpunct}{\mcitedefaultseppunct}\relax
\EndOfBibitem
\bibitem{Chekanov:2004mw}
{\bf ZEUS} collaboration, S.~Chekanov {\em et.~al.},  \mbox{}
  \href{http://dx.doi.org/10.1016/j.nuclphysb.2004.06.034}{{\em Nucl. Phys.}
  {\bf B695} (2004) 3} [\href{http://arXiv.org/abs/hep-ex/0404008}{{\tt
  arXiv:hep-ex/0404008}}]\relax
\mciteBstWouldAddEndPuncttrue
\mciteSetBstMidEndSepPunct{\mcitedefaultmidpunct}
{\mcitedefaultendpunct}{\mcitedefaultseppunct}\relax
\EndOfBibitem
\bibitem{Aktas:2005xu}
{\bf H1} collaboration, A.~Aktas {\em et.~al.},  \mbox{}
  \href{http://dx.doi.org/10.1140/epjc/s2006-02519-5}{{\em Eur. Phys. J.} {\bf
  C46} (2006) 585} [\href{http://arXiv.org/abs/hep-ex/0510016}{{\tt
  arXiv:hep-ex/0510016}}]\relax
\mciteBstWouldAddEndPuncttrue
\mciteSetBstMidEndSepPunct{\mcitedefaultmidpunct}
{\mcitedefaultendpunct}{\mcitedefaultseppunct}\relax
\EndOfBibitem
\bibitem{McLerran:1994ni}
L.~D. McLerran and R.~Venugopalan,  \mbox{}
  \href{http://dx.doi.org/10.1103/PhysRevD.49.2233}{{\em Phys. Rev.} {\bf D49}
  (1994) 2233} [\href{http://arXiv.org/abs/hep-ph/9309289}{{\tt
  arXiv:hep-ph/9309289}}]\relax
\mciteBstWouldAddEndPuncttrue
\mciteSetBstMidEndSepPunct{\mcitedefaultmidpunct}
{\mcitedefaultendpunct}{\mcitedefaultseppunct}\relax
\EndOfBibitem
\bibitem{Kuokkanen:2011je}
J.~Kuokkanen, K.~Rummukainen and H.~Weigert,  \mbox{}
  \href{http://arXiv.org/abs/1108.1867}{{\tt arXiv:1108.1867 [hep-ph]}}\relax
\mciteBstWouldAddEndPuncttrue
\mciteSetBstMidEndSepPunct{\mcitedefaultmidpunct}
{\mcitedefaultendpunct}{\mcitedefaultseppunct}\relax
\EndOfBibitem
\bibitem{Kovchegov:2006vj}
Y.~V. Kovchegov and H.~Weigert,  \mbox{}
  \href{http://dx.doi.org/10.1016/j.nuclphysa.2006.10.075}{{\em Nucl. Phys.}
  {\bf A784} (2007) 188} [\href{http://arXiv.org/abs/hep-ph/0609090}{{\tt
  arXiv:hep-ph/0609090 [hep-ph]}}]\relax
\mciteBstWouldAddEndPuncttrue
\mciteSetBstMidEndSepPunct{\mcitedefaultmidpunct}
{\mcitedefaultendpunct}{\mcitedefaultseppunct}\relax
\EndOfBibitem
\bibitem{Lappi:2012vw}
T.~Lappi and H.~M{\"a}ntysaari,  \mbox{}
  \href{http://dx.doi.org/10.1140/epjc/s10052-013-2307-z}{{\em Eur. Phys. J.}
  {\bf C73} (2013) 2307} [\href{http://arXiv.org/abs/1212.4825}{{\tt
  arXiv:1212.4825 [hep-ph]}}]\relax
\mciteBstWouldAddEndPuncttrue
\mciteSetBstMidEndSepPunct{\mcitedefaultmidpunct}
{\mcitedefaultendpunct}{\mcitedefaultseppunct}\relax
\EndOfBibitem
\bibitem{Kovchegov:2001sc}
Y.~V. Kovchegov and K.~Tuchin,  \mbox{}
  \href{http://dx.doi.org/10.1103/PhysRevD.65.074026}{{\em Phys. Rev.} {\bf
  D65} (2002) 074026} [\href{http://arXiv.org/abs/hep-ph/0111362}{{\tt
  arXiv:hep-ph/0111362 [hep-ph]}}]\relax
\mciteBstWouldAddEndPuncttrue
\mciteSetBstMidEndSepPunct{\mcitedefaultmidpunct}
{\mcitedefaultendpunct}{\mcitedefaultseppunct}\relax
\EndOfBibitem
\bibitem{Blaizot:2010kh}
J.~P. Blaizot, T.~Lappi and Y.~Mehtar-Tani,  \mbox{}
  \href{http://dx.doi.org/10.1016/j.nuclphysa.2010.06.009}{{\em Nucl. Phys.}
  {\bf A846} (2010) 63} [\href{http://arXiv.org/abs/1005.0955}{{\tt
  arXiv:1005.0955 [hep-ph]}}]\relax
\mciteBstWouldAddEndPuncttrue
\mciteSetBstMidEndSepPunct{\mcitedefaultmidpunct}
{\mcitedefaultendpunct}{\mcitedefaultseppunct}\relax
\EndOfBibitem
\bibitem{Kharzeev:2003wz}
D.~Kharzeev, Y.~V. Kovchegov and K.~Tuchin,  \mbox{}
  \href{http://dx.doi.org/10.1103/PhysRevD.68.094013}{{\em Phys. Rev.} {\bf
  D68} (2003) 094013} [\href{http://arXiv.org/abs/hep-ph/0307037}{{\tt
  arXiv:hep-ph/0307037 [hep-ph]}}]\relax
\mciteBstWouldAddEndPuncttrue
\mciteSetBstMidEndSepPunct{\mcitedefaultmidpunct}
{\mcitedefaultendpunct}{\mcitedefaultseppunct}\relax
\EndOfBibitem
\bibitem{Blaizot:2004wu}
J.~P. Blaizot, F.~Gelis and R.~Venugopalan,  \mbox{}
  \href{http://dx.doi.org/10.1016/j.nuclphysa.2004.07.005}{{\em Nucl. Phys.}
  {\bf A743} (2004) 13} [\href{http://arXiv.org/abs/hep-ph/0402256}{{\tt
  arXiv:hep-ph/0402256 [hep-ph]}}]\relax
\mciteBstWouldAddEndPuncttrue
\mciteSetBstMidEndSepPunct{\mcitedefaultmidpunct}
{\mcitedefaultendpunct}{\mcitedefaultseppunct}\relax
\EndOfBibitem
\bibitem{Dominguez:2011wm}
F.~Dominguez, C.~Marquet, B.-W. Xiao and F.~Yuan,  \mbox{}
  \href{http://dx.doi.org/10.1103/PhysRevD.83.105005}{{\em Phys. Rev.} {\bf
  D83} (2011) 105005} [\href{http://arXiv.org/abs/1101.0715}{{\tt
  arXiv:1101.0715 [hep-ph]}}]\relax
\mciteBstWouldAddEndPuncttrue
\mciteSetBstMidEndSepPunct{\mcitedefaultmidpunct}
{\mcitedefaultendpunct}{\mcitedefaultseppunct}\relax
\EndOfBibitem
\bibitem{Pumplin:2002vw}
J.~Pumplin, D.~Stump, J.~Huston, H.~Lai, P.~M. Nadolsky {\em et.~al.},  \mbox{}
   {\em JHEP} {\bf 0207} (2002) 012
  [\href{http://arXiv.org/abs/hep-ph/0201195}{{\tt arXiv:hep-ph/0201195
  [hep-ph]}}]\relax
\mciteBstWouldAddEndPuncttrue
\mciteSetBstMidEndSepPunct{\mcitedefaultmidpunct}
{\mcitedefaultendpunct}{\mcitedefaultseppunct}\relax
\EndOfBibitem
\bibitem{Aaron:2009xp}
{\bf H1} collaboration, F.~Aaron {\em et.~al.},  \mbox{}
  \href{http://dx.doi.org/10.1007/JHEP05(2010)032}{{\em JHEP} {\bf 1005} (2010)
  032} [\href{http://arXiv.org/abs/0910.5831}{{\tt arXiv:0910.5831
  [hep-ex]}}]\relax
\mciteBstWouldAddEndPuncttrue
\mciteSetBstMidEndSepPunct{\mcitedefaultmidpunct}
{\mcitedefaultendpunct}{\mcitedefaultseppunct}\relax
\EndOfBibitem
\bibitem{Frankfurt:2010ea}
L.~Frankfurt, M.~Strikman and C.~Weiss,  \mbox{}
  \href{http://dx.doi.org/10.1103/PhysRevD.83.054012}{{\em Phys. Rev.} {\bf
  D83} (2011) 054012} [\href{http://arXiv.org/abs/1009.2559}{{\tt
  arXiv:1009.2559 [hep-ph]}}]\relax
\mciteBstWouldAddEndPuncttrue
\mciteSetBstMidEndSepPunct{\mcitedefaultmidpunct}
{\mcitedefaultendpunct}{\mcitedefaultseppunct}\relax
\EndOfBibitem
\bibitem{Adams:2006uz}
{\bf STAR} collaboration, J.~Adams {\em et.~al.},  \mbox{}
  \href{http://dx.doi.org/10.1103/PhysRevLett.97.152302}{{\em Phys. Rev. Lett.}
  {\bf 97} (2006) 152302} [\href{http://arXiv.org/abs/nucl-ex/0602011}{{\tt
  arXiv:nucl-ex/0602011 [nucl-ex]}}]\relax
\mciteBstWouldAddEndPuncttrue
\mciteSetBstMidEndSepPunct{\mcitedefaultmidpunct}
{\mcitedefaultendpunct}{\mcitedefaultseppunct}\relax
\EndOfBibitem
\bibitem{Aaltonen:2009ne}
{\bf CDF} collaboration, T.~Aaltonen {\em et.~al.},  \mbox{}
  \href{http://dx.doi.org/10.1103/PhysRevD.82.119903,
  10.1103/PhysRevD.79.112005}{{\em Phys. Rev.} {\bf D79} (2009) 112005}
  [\href{http://arXiv.org/abs/0904.1098}{{\tt arXiv:0904.1098 [hep-ex]}}]\relax
\mciteBstWouldAddEndPuncttrue
\mciteSetBstMidEndSepPunct{\mcitedefaultmidpunct}
{\mcitedefaultendpunct}{\mcitedefaultseppunct}\relax
\EndOfBibitem
\bibitem{Antchev:2011vs}
{\bf TOTEM} collaboration, G.~Antchev, P.~Aspell, I.~Atanassov, V.~Avati,
  J.~Baechler {\em et.~al.},  \mbox{}
  \href{http://dx.doi.org/10.1209/0295-5075/96/21002}{{\em Europhys. Lett.}
  {\bf 96} (2011) 21002} [\href{http://arXiv.org/abs/1110.1395}{{\tt
  arXiv:1110.1395 [hep-ex]}}]\relax
\mciteBstWouldAddEndPuncttrue
\mciteSetBstMidEndSepPunct{\mcitedefaultmidpunct}
{\mcitedefaultendpunct}{\mcitedefaultseppunct}\relax
\EndOfBibitem
\bibitem{Tribedy:2010ab}
P.~Tribedy and R.~Venugopalan,  \mbox{}
  \href{http://dx.doi.org/10.1016/j.nuclphysa.2011.04.008,
  10.1016/j.nuclphysa.2010.12.006}{{\em Nucl. Phys.} {\bf A850} (2011) 136}
  [\href{http://arXiv.org/abs/1011.1895}{{\tt arXiv:1011.1895 [hep-ph]}}]\relax
\mciteBstWouldAddEndPuncttrue
\mciteSetBstMidEndSepPunct{\mcitedefaultmidpunct}
{\mcitedefaultendpunct}{\mcitedefaultseppunct}\relax
\EndOfBibitem
\bibitem{deFlorian:2007aj}
D.~de~Florian, R.~Sassot and M.~Stratmann,  \mbox{}
  \href{http://dx.doi.org/10.1103/PhysRevD.75.114010}{{\em Phys.Rev.} {\bf D75}
  (2007) 114010} [\href{http://arXiv.org/abs/hep-ph/0703242}{{\tt
  arXiv:hep-ph/0703242 [HEP-PH]}}]\relax
\mciteBstWouldAddEndPuncttrue
\mciteSetBstMidEndSepPunct{\mcitedefaultmidpunct}
{\mcitedefaultendpunct}{\mcitedefaultseppunct}\relax
\EndOfBibitem
\bibitem{Kowalski:2007rw}
H.~Kowalski, T.~Lappi and R.~Venugopalan,  \mbox{}
  \href{http://dx.doi.org/10.1103/PhysRevLett.100.022303}{{\em Phys. Rev.
  Lett.} {\bf 100} (2008) 022303} [\href{http://arXiv.org/abs/0705.3047}{{\tt
  arXiv:0705.3047 [hep-ph]}}]\relax
\mciteBstWouldAddEndPuncttrue
\mciteSetBstMidEndSepPunct{\mcitedefaultmidpunct}
{\mcitedefaultendpunct}{\mcitedefaultseppunct}\relax
\EndOfBibitem
\bibitem{GolecBiernat:2003ym}
K.~J. Golec-Biernat and A.~Stasto,  \mbox{}
  \href{http://dx.doi.org/10.1016/j.nuclphysb.2003.07.011}{{\em Nucl. Phys.}
  {\bf B668} (2003) 345} [\href{http://arXiv.org/abs/hep-ph/0306279}{{\tt
  arXiv:hep-ph/0306279 [hep-ph]}}]\relax
\mciteBstWouldAddEndPuncttrue
\mciteSetBstMidEndSepPunct{\mcitedefaultmidpunct}
{\mcitedefaultendpunct}{\mcitedefaultseppunct}\relax
\EndOfBibitem
\bibitem{Berger:2010sh}
J.~Berger and A.~Stasto,  \mbox{}
  \href{http://dx.doi.org/10.1103/PhysRevD.83.034015}{{\em Phys. Rev.} {\bf
  D83} (2011) 034015} [\href{http://arXiv.org/abs/1010.0671}{{\tt
  arXiv:1010.0671 [hep-ph]}}]\relax
\mciteBstWouldAddEndPuncttrue
\mciteSetBstMidEndSepPunct{\mcitedefaultmidpunct}
{\mcitedefaultendpunct}{\mcitedefaultseppunct}\relax
\EndOfBibitem
\bibitem{Berger:2011ew}
J.~Berger and A.~M. Stasto,  \mbox{}
  \href{http://dx.doi.org/10.1103/PhysRevD.84.094022}{{\em Phys. Rev.} {\bf
  D84} (2011) 094022} [\href{http://arXiv.org/abs/1106.5740}{{\tt
  arXiv:1106.5740 [hep-ph]}}]\relax
\mciteBstWouldAddEndPuncttrue
\mciteSetBstMidEndSepPunct{\mcitedefaultmidpunct}
{\mcitedefaultendpunct}{\mcitedefaultseppunct}\relax
\EndOfBibitem
\bibitem{Berger:2012wx}
J.~Berger and A.~M. Stasto,  \mbox{}
  \href{http://arXiv.org/abs/1205.2037}{{\tt arXiv:1205.2037 [hep-ph]}}\relax
\mciteBstWouldAddEndPuncttrue
\mciteSetBstMidEndSepPunct{\mcitedefaultmidpunct}
{\mcitedefaultendpunct}{\mcitedefaultseppunct}\relax
\EndOfBibitem
\bibitem{Albacete:2010bs}
J.~L. Albacete and C.~Marquet,  \mbox{}
  \href{http://dx.doi.org/10.1016/j.physletb.2010.02.073}{{\em Phys. Lett.}
  {\bf B687} (2010) 174} [\href{http://arXiv.org/abs/1001.1378}{{\tt
  arXiv:1001.1378 [hep-ph]}}]\relax
\mciteBstWouldAddEndPuncttrue
\mciteSetBstMidEndSepPunct{\mcitedefaultmidpunct}
{\mcitedefaultendpunct}{\mcitedefaultseppunct}\relax
\EndOfBibitem
\bibitem{Adare:2011sc}
{\bf PHENIX} collaboration, A.~Adare {\em et.~al.},  \mbox{}
  \href{http://dx.doi.org/10.1103/PhysRevLett.107.172301}{{\em Phys. Rev.
  Lett.} {\bf 107} (2011) 172301} [\href{http://arXiv.org/abs/1105.5112}{{\tt
  arXiv:1105.5112 [nucl-ex]}}]\relax
\mciteBstWouldAddEndPuncttrue
\mciteSetBstMidEndSepPunct{\mcitedefaultmidpunct}
{\mcitedefaultendpunct}{\mcitedefaultseppunct}\relax
\EndOfBibitem
\bibitem{Arsene:2004ux}
{\bf BRAHMS} collaboration, I.~Arsene {\em et.~al.},  \mbox{}
  \href{http://dx.doi.org/10.1103/PhysRevLett.93.242303}{{\em Phys. Rev. Lett.}
  {\bf 93} (2004) 242303} [\href{http://arXiv.org/abs/nucl-ex/0403005}{{\tt
  arXiv:nucl-ex/0403005 [nucl-ex]}}]\relax
\mciteBstWouldAddEndPuncttrue
\mciteSetBstMidEndSepPunct{\mcitedefaultmidpunct}
{\mcitedefaultendpunct}{\mcitedefaultseppunct}\relax
\EndOfBibitem
\bibitem{Abelev:2012cn}
{\bf ALICE} collaboration, B.~Abelev {\em et.~al.},  \mbox{}
  \href{http://dx.doi.org/10.1016/j.physletb.2012.09.015}{{\em Phys. Lett.}
  {\bf B717} (2012) 162} [\href{http://arXiv.org/abs/1205.5724}{{\tt
  arXiv:1205.5724 [hep-ex]}}]\relax
\mciteBstWouldAddEndPuncttrue
\mciteSetBstMidEndSepPunct{\mcitedefaultmidpunct}
{\mcitedefaultendpunct}{\mcitedefaultseppunct}\relax
\EndOfBibitem
\bibitem{Khachatryan:2010us}
{\bf CMS} collaboration, V.~Khachatryan {\em et.~al.},  \mbox{}
  \href{http://dx.doi.org/10.1103/PhysRevLett.105.022002}{{\em Phys. Rev.
  Lett.} {\bf 105} (2010) 022002} [\href{http://arXiv.org/abs/1005.3299}{{\tt
  arXiv:1005.3299 [hep-ex]}}]\relax
\mciteBstWouldAddEndPuncttrue
\mciteSetBstMidEndSepPunct{\mcitedefaultmidpunct}
{\mcitedefaultendpunct}{\mcitedefaultseppunct}\relax
\EndOfBibitem
\bibitem{Meredith:2011nla}
B.~A. Meredith, {\em {A Study of Nuclear Effects using Forward-Rapidity Hadron
  Production and Di-Hadron Angular Correlations in $\sqrt{s_{NN}} =200$ GeV
  d+Au and p+p Collisions with the PHENIX Detector at RHIC}}.
\newblock {Ph.D. thesis at University of Illinois at Urbana-Champaign},
  2011\relax
\mciteBstWouldAddEndPuncttrue
\mciteSetBstMidEndSepPunct{\mcitedefaultmidpunct}
{\mcitedefaultendpunct}{\mcitedefaultseppunct}\relax
\EndOfBibitem
\bibitem{Abelev:2012mj}
{\bf ALICE} collaboration, B.~Abelev {\em et.~al.},  \mbox{}
  \href{http://arXiv.org/abs/1210.4520}{{\tt arXiv:1210.4520 [nucl-ex]}}\relax
\mciteBstWouldAddEndPuncttrue
\mciteSetBstMidEndSepPunct{\mcitedefaultmidpunct}
{\mcitedefaultendpunct}{\mcitedefaultseppunct}\relax
\EndOfBibitem
\bibitem{Eskola:2002kv}
K.~Eskola and H.~Honkanen,  \mbox{}
  \href{http://dx.doi.org/10.1016/S0375-9474(02)01304-0}{{\em Nucl. Phys.} {\bf
  A713} (2003) 167} [\href{http://arXiv.org/abs/hep-ph/0205048}{{\tt
  arXiv:hep-ph/0205048 [hep-ph]}}]\relax
\mciteBstWouldAddEndPuncttrue
\mciteSetBstMidEndSepPunct{\mcitedefaultmidpunct}
{\mcitedefaultendpunct}{\mcitedefaultseppunct}\relax
\EndOfBibitem
\bibitem{Helenius:2012wd}
I.~Helenius, K.~J. Eskola, H.~Honkanen and C.~A. Salgado,  \mbox{}
  \href{http://dx.doi.org/10.1007/JHEP07(2012)073}{{\em JHEP} {\bf 1207} (2012)
  073} [\href{http://arXiv.org/abs/1205.5359}{{\tt arXiv:1205.5359
  [hep-ph]}}]\relax
\mciteBstWouldAddEndPuncttrue
\mciteSetBstMidEndSepPunct{\mcitedefaultmidpunct}
{\mcitedefaultendpunct}{\mcitedefaultseppunct}\relax
\EndOfBibitem
\bibitem{helenius2013rapidity}
I.~Helenius.
\newblock Private communication\relax
\mciteBstWouldAddEndPuncttrue
\mciteSetBstMidEndSepPunct{\mcitedefaultmidpunct}
{\mcitedefaultendpunct}{\mcitedefaultseppunct}\relax
\EndOfBibitem
\end{mcitethebibliography}\endgroup
\bibliographystyle{JHEP-2modM}

\end{document}